\documentclass[12pt,a4paper]{article}

\usepackage[latin1,utf8]{inputenc}
\usepackage{amsfonts,amsmath,amstext,amsbsy,amscd,amssymb}
\usepackage{graphicx}
\usepackage[authoryear,square]{natbib}
\usepackage{float}
\usepackage{placeins}
\usepackage{mathptmx}
\usepackage{color}

\def\R{\mathbb{R}}
\def\Z{\mathbb{Z}}

\def\P{\mathbb{P}}

\makeatletter    

\newcommand{\unnumberedcaption}%
	{\@dblarg{\@unnumberedcaption\@captype}}

\newcommand{\@unnumberedcaption}{}
\long\def\@unnumberedcaption#1[#2]#3{\par
  \addcontentsline{\csname ext@#1\endcsname}{#1}{%
    \protect\numberline{}{\ignorespaces #2}%
    }%
  \begingroup
    \@parboxrestore
    \normalsize
    \@makeunnumberedcaption{\ignorespaces #3}\par
  \endgroup}

\newcommand{\@makeunnumberedcaption}[1]{%
  \vskip\abovecaptionskip
  \sbox\@tempboxa{#1}%
  \ifdim \wd\@tempboxa >\hsize
    #1\par
  \else
    \global \@minipagefalse
    \hbox to\hsize{\hfil\box\@tempboxa\hfil}%
  \fi
  \vskip\belowcaptionskip}

\@ifundefined{abovecaptionskip}{%
  \newlength{\abovecaptionskip}%
  \setlength{\abovecaptionskip}{10pt}%
}{}
\@ifundefined{belowcaptionskip}{%
  \newlength{\belowcaptionskip}%
  \setlength{\belowcaptionskip}{0pt}%
}{}

\makeatother    

\newtheorem{rem}{\it Remark}

\newcommand{\som}[2]{\displaystyle{\sum_{#1}^{#2}}}

\newcommand{\trans}[1]{\,{\vphantom{#1}}^{t}\!{#1}}

\newcommand{\tend}[2]{\underset{{#1}\rightarrow {#2}}{\longrightarrow}}

\newcommand{\integ}[2]{\displaystyle{\int_{#1}^{#2}}}

\newcommand{\norm}[1]{\lVert #1 \rVert}
\newcommand{\deriv}[2]{\frac{\partial {#1}}{\partial {#2}}}

\newcommand{\super}[1]{\underset{#1}{sup}}

\newcommand{\ev}{\textcolor{black}}
\newcommand{\cl}{\textcolor{black}}

\begin{document}

\title{{\bf Approximation of epidemic models by diffusion processes and their statistical inference}}
\author{Romain  Guy$^{1,2,*}$\and  Catherine Lar\'edo$^{1,2}$ \and Elisabeta Vergu$^1$}
\maketitle
{\sc \it \noindent
$^{1}$ UR 341 Math\'ematiques et Informatique Appliqu\'ees, INRA, Jouy-en-Josas, France \\
$^{2}$ UMR 7599 Laboratoire de Probabilit\'es et Mod\`eles al\'eatoires, Universit\'e Denis Diderot Paris 7 and CNRS, Paris, France \\
$^*$Corresponding author; e-mail: romain.guy@jouy.inra.fr}

\begin{abstract}
Multidimensional continuous-time Markov jump processes \ev{$(Z(t))$} on $\Z^p$ form a usual set-up for modeling $SIR$-like epidemics. However, when facing incomplete epidemic data, inference based on \ev{$(Z(t))$} is not easy to be achieved. Here, we start building a new framework for the estimation of key parameters of epidemic models based on statistics of diffusion processes approximating \ev{$(Z(t))$}. First, \ev{previous results} on the approximation of density-dependent $SIR$-like models by diffusion processes with small diffusion coefficient $\frac{1}{\sqrt{N}}$, where $N$ is the population size, are generalized to non-autonomous systems. Second, \ev{our previous inference results} on discretely observed diffusion processes with small diffusion coefficient are extended to time-dependent diffusions. Consistent and asymptotically Gaussian estimates are obtained for a fixed number $n$ of observations, which corresponds to the epidemic context, and for $N\rightarrow \infty$. A correction term, which yields better estimates non asymptotically, is also included. Finally, performances and robustness of our estimators with respect to various parameters such as $R_0$ (the basic reproduction number), $N$, $n$ are investigated on simulations. Two models, $SIR$ and $SIRS$, corresponding to single and recurrent outbreaks, respectively, are used to \ev{simulate data. The} findings indicate that our estimators have good asymptotic properties and behave noticeably well for realistic numbers of observations and population sizes. This study lays the foundations of a generic inference method currently under extension to incompletely observed epidemic data. Indeed, contrary to the majority of current inference techniques for partially observed processes, which necessitates computer intensive simulations, our method being mostly an analytical approach requires only the classical optimization steps.

\end{abstract}	

\section{Introduction}
\label{intro}
Mathematical modeling of epidemic spread and estimation of key parameters from data provided much insight in the understanding of public health problems related to infectious diseases. Classically, an epidemic dynamics in closed population of size $N$ is described by the $SIR$ model (Susceptible-Infectious-Removed from
the infectious chain), where each individual can find himself at a given time in one of these three mutually exclusive health states. Systems with larger dimensionality can be obtained if the description of the epidemic dynamics is refined (\citet{kee11}). One of the most natural representations of the $SIR$ model is a bidimensional continuous-time Markov jump process. \ev{Given that $R(t)=N-S(t)-I(t)$ $\forall t$, we can define $Z (t)=(S(t),I(t))$ with initial state $Z (0)=(S_0,I_0)$ and transitions }
$(S,I)\stackrel{\frac{\lambda}{N}SI}{\longrightarrow}(S-1,I+1)$ and
$(S,I)\stackrel{\gamma I}{\longrightarrow}(S,I-1)$, where $\lambda$ is the transmission rate and $\gamma=1/d$ is the recovery rate or the inverse of the mean infection duration $d$. Beyond this process, the successive transitions of individuals between states were described in various mathematical frameworks: \ev{ODE/PDE (\citet{die00})\cl{;} difference equations and continuous or discrete-time stochastic processes (\citet{dal01,die12}),  such as point processes, renewal processes, branching processes, diffusion processes.}\\
These models are naturally parametric and allow estimating key parameters (such as transmission rate, mean sojourn time in the infectious state, extinction probability) through likelihood-based or M-estimation methods sometimes coupled to Bayesian methods \cl{(\citet{and00})}. Due to the fact that epidemic data are most often partially observed (e.g. infectious and recovery dates are not observed for all individuals recorded in the surveillance system, not all the infectious individuals are reported) and also temporally and/or  spatially aggregated, estimation through likelihood-based approaches is rarely straightforward, whatever the mathematical representation used. Various methods
were developed during the last years to overcome this problem, data augmentation methods and likelihood-free methods being those which generated the keenest interest (\citet{bre09,mck09}). Although these methods allow considering different patterns of missingness, they do not provide a definitive solution to the  statistical inference from epidemic data. \ev{Indeed, in practice there are some limitations due to the amount of augmented data and to the adjustment of the numerous tuning parameters} (\citet{and00}; see also \citet{one10} for a short review of available statistical methods for relating models to data and for future challenges). Moreover, identifiability related issues are rarely addressed.\\
In this context, diffusion processes, which provide good approximations of epidemic dynamics \cl{(see e.g. \citet{fuc13, ros09})}, allow shedding new light on inference related problems of epidemic data due to their analytical power. The normalization of the $SIR$ Markov jump process \ev{$\cl{(Z(t))}$ by $N$ asymptotically leads to an ODE system $x(t)=(s(t),i(t))$ with $s(0)=1-I(0)/N, i(0)= I(0)/N$ and $ds(t)/dt=-\lambda s(t)i(t)$ and $di(t)/dt=\lambda s(t)i(t)-\gamma i(t)$.} Before passing to the limit, one can describe the epidemic dynamics through a bidimensional diffusion with a small diffusion coefficient proportional to $1/\sqrt{N}$. On the statistical side, we can consider as a first approximation that epidemic data correspond to low frequency data (i.e. a fixed number of observations $n$) observed on a fixed interval $[0,T]$. Although this is an optimistic view of field data which are most often incomplete and incorporate observational errors, it lays the foundations of further investigations in more realistic contexts. \\
Historically, statistics for diffusions were developed for continuously observed processes which renders possible getting an explicit formulation of the likelihood (\citet{kut84,lip01}). As mentioned above, in practice, epidemic data are not continuous, but partial, with various mechanisms underlying the missingness and leading to intractable likelihoods: trajectories can be discretely observed with a sampling interval (low frequency or high frequency observations, i.e. $n \rightarrow \infty$); discrete observations can correspond to integrated processes; some coordinates can be unobserved. Since the nineties, statistical methods associated to the first two types of data have been developed (e.g. \citet{gen93,kes00,gen00,glo01}). Recently proposed approaches for multidimensional diffusions are based on the filtering theory (\citet{gen06,fea08}). Concerning diffusions with small diffusion coefficient from discrete observations, the asymptotic properties of estimators were largely studied over the two last decades (e.g. \citet{lar90,gen90,gen02,sor03,glo09}) in various contexts (uni- and multidimensional cases, observations sampled at low and high frequency, discrete sampling of the state space). In this framework, it is important to distinguish drift parameters and diffusion parameters, because they are not estimated at the same rate. In a recent work (\cite{guy12}), we studied multidimensional diffusion with small variance, for discrete  observations on both components of the system approximating \cl{$(Z(t))$}. We provided minimum contrast estimators with good properties for both high and low-frequency observations on a fixed time interval $[0,T]$.\\
In this paper, we propose first, based on the results of \cite{eth05}, a generic and rigorous method to construct multidimensional diffusion processes with small variance as mathematical representations of epidemic dynamics, by approximating a Markov jump process (Section \ref{diff_approx}). The approach of \cite{eth05} is extended to general density time-dependent Markov processes (Section \ref{time_dep}). The second and main result is a new inference method for the parameters of the diffusion process obtained in Section \ref{diff_approx}, discretely observed (fixed $n$) on a fixed time interval and for the special case where the same parameters are in the diffusion and drift terms. Building on the results of \cite{guy12}, we elaborate a new contrast based on the Gaussian approximation of the diffusion process (Section \ref{infer}). In addition to consistent and asymptotically Gaussian minimum contrast estimators obtained for fixed $n$ (which corresponds to the epidemic context) and for $N\rightarrow \infty$, the correction term we introduce in the new contrast allows yielding better estimates non asymptotically. We also extent the results of \cite{guy12} to time-dependent diffusions. Finally, the accuracy of these estimators is explored on simulated  epidemic data (Section \ref{simul}) for single outbreaks ($SIR$) and for recurrent epidemics (non autonomous $SIRS$, i.e. with seasonal forcing in transmission). \ev{The case of discrete observations of all the coordinates that we investigate in this study also provides a best case scenario to correctly assess the performances of the inference method for incomplete (in time and state space) data in further research.} More generally, our study lays the foundations of the inference approach based on partially observed integrated diffusions that we are currently investigating \ev{(short discussion in Section \ref{conc})}.


\section{Construction of the diffusion approximation of epidemic models}
\label{diff_approx}
In this section we present the generic procedure for building the  diffusion approximation of \ev{a density
dependent jump Markov process $(Z(t))$} in ${ \Z}^p$ as proposed in
\citet{eth05}. Then, we derive it for two $SIR$-like epidemic
models and extend it to general density time-dependent Markov processes. A first normalization, corresponding to a
law of large numbers, provides the convergence of \cl{$(Z(t))$} to a deterministic limit $x(t)$, solution of an ordinary differential equation. Then, centering \cl{$(Z(t))$}, a central limit theorem yields that the process $\sqrt N (\frac{Z (t)}{N}-x(t))$ is approximated either by a Gaussian process (\citet{kam92}) or by a diffusion process, these two approximations being essentially equivalent processes at least on fixed time intervals (see \citet{eth05} Chapter 11, Section 2\cl{.}3).\\
In fact, the diffusion approximation possesses a small diffusion coefficient
proportional to $N^{-1/2}$ and the Gaussian process comes from large deviations
and corresponds to the first two terms of the diffusion expansion (see
\citet{aze82,fre84}). We chose here the diffusion approximation, as our
theoretical results supporting the estimation of epidemic model parameters are
built on it (but they are still valid for the Gaussian
approximation of the Markov jump process).

\subsection{A different representation of Markov jump processes \cl{(Ethier and Kurtz, 2005)}}
\label{Representation}
A  multidimensional Markov jump process  $(Z(t), t \geq 0)$ with state space $ E \subset  \Z^p$ is usually  described by an initial distribution $\lambda(.)$ on
$E$, and  a collection of non negative functions $(\alpha_l (.): E \rightarrow \R^+)$  indexed by $l \in \Z^p , l \neq (0,\dots,0)$ that satisfy,
\begin{equation}\label{H:bound_alpha}\forall k \in E,\;\;  0 <\som{l\in\Z^p}{}
\alpha_l(k)
 =\alpha(k)< \infty.\end{equation}\\
These functions represent the transition intensities of \cl{$(Z(t))$} by setting the transition rates from  $k$ to $k+l$,
\begin{equation}\label{trans}
 q_{k, k+l}=  \alpha_l(k).
\end{equation}\\
The collection $(\alpha_l(k))_l$ is associated to all the possible jumps from state $k$, and the time spent in this state is exponentially distributed \cl{with parameter} $ \alpha(k)$. \ev{The transition probabilities of the embedded Markov chain $(C_m)$ are, for $k$ and $k+l \in E$, ${\mathbb P}(C_{m+1}=k+l|C_m=k)=\alpha_{l}(k)/{\alpha(k)}$.}\\
\cl{The generator $\cal A$} of the Markov jump process $(Z(t), t\geq 0)$  \cl{ is defined on the set of}  real measurable and bounded functions $f$ on $(E,{\cal B}(E))$ and \cl{writes,} 
 \begin{equation}\label{GenZ}
\cl{\cal A}f(k) =\sum_{l \in \Z^p} \alpha_l(k)(f(k+l)-f(k))=\alpha(k)\sum_{l \in \Z^p}
(f(k+l)-f(k))\frac{ \alpha_l(k)}{\alpha(k)}.
\end{equation}\\
Following \cl{Chapter 6 Section 4 of} \citet{eth05}, there is  another useful expression based on Poisson processes for $Z(t)$. 
Let  $ (P_l(.))$ be a family of independent Poisson processes with rate 1, indexed by $l\in \Z^p$, independent of $Z(0)$. After applying to each $P_l(.)$ a random time change based on $\alpha_l(.)$, $Z(t)$ can be expressed as,
\begin{equation}\label{Poisson}
Z(t)= Z(0) +\sum_{l \in \Z^p} l\; P_l\biggl(\int_0^t \alpha_l(Z(u))du\biggr).
\end{equation}
This new expression of the jump process $Z(t)$, obtained by proving the equality of the two infinitesimal generators associated to each representation\cl{,} is very powerful to evaluate  distances between trajectories of processes, and consequently to establish approximation results. All technical details are provided in Appendix \ref{App:Poisson_repr}.

\begin{rem}
\ev{From the point of view of simulation algorithms for epidemic dynamics, the classical representation based on transition rates \eqref{trans} relates to the algorithm of \citet{gil77}, }
whereas the time changed Poisson representation \eqref{Poisson} corresponds to the more general scheme of \citet{sel83} (typically for non exponential distributions of the infectious period). 
\end{rem}

\subsection{The generic approach for the diffusion approximation}
\label{generic}
Let us now consider density dependent Markov jump processes on $\Z^p$. \cl{The Markov process  $Z(t)$ has} state space 
$E= \{0,\dots,N \}^p $, \ev{where $N$ (fixed parameter) represents the total population size,} transition rates  $q_{k,k+l} \;=\; \alpha_{l}(k)$ and jumps \cl{$l$} in
$E^-=\{-N,\dots,N\}^p $ \cl{such that $k,k+l \in E, l\neq \{0\}^p$}.\\ 
For \cl{$ l \in \Z^p$ or  $y \in {\mathbb R}^p$ with components $ y_i$, denote by $\trans{l}$ and $\trans{y}$ the transpositions of $l$ and $y$ respectively, and by} $[y]$ the vector
of ${\mathbb Z}^p$ with components $[y_i]$,  where $[y_i]$ is the integer part of $y_i$.\\ In order to describe the behavior of $Z(t)$ for large $N$, we
assume,\\

\noindent 
{\bf (H1)}: \cl{$\forall l \in E^-,\;\forall y \in  [0,1]^p\;\;$}  
$\frac{1}{N} \alpha_l([Ny]) \tend{N}{+\infty}\beta_l(y)$,\\ 
{\bf (H2)}: $\forall l\in E^-, \; \beta_l\in C^2([0,1]^p).$\\

\noindent
These assumptions ensure that the Markov jump process is a density dependent
process (H1) with asymptotic regularity properties for its transition
rates (H2). Note that for density dependent processes, the collection of
functions $\alpha_l$ and $\beta_l$ is finite \cl{and the condition $\alpha(k)< \infty$ 
of \eqref{H:bound_alpha} is satisfied;} for more general processes, additional assumptions on $\beta_l$
similar to \eqref{H:bound_alpha} are required.

\noindent
From the original jump process \cl{$(Z(t))$} on $E= \{0,\dots, N\}^p$, we consider the normalized jump Markov process \cl{$(Z_N(t) =\frac{Z(t)}{N})$}  on $E_N=\{N^{-1}k, \ k \in E\}$. It satisfies, using \eqref{trans}, \cl{$Z_N(0) =\frac{Z(0)}{N}$ and for $y, y+z \in E_N$},
\begin{equation}\label{barZN}
 \cl{q^{N}_{y,y+z}= \alpha_{Nz}(Ny) \;;\; 
	{\cal A}_N f(y)= \sum_{l \in E^-}  \alpha_{l}(Ny)(f(y+\frac{l}{N})-f(y)).}
\end{equation}
The time changed Poisson process representation
\eqref{Poisson} is  
\begin{equation}\label{PoissonN}
	Z_N(t)= Z_N(0) +\sum_{l \in E^{-}} \frac{l}{N}\; P_l\biggl(\int_0^t \alpha_l(NZ_N(u))du\biggr).
\end{equation}
In order to build approximation processes from \cl{$(Z_N(t))$}, the first step is to assess the mean behavior of \cl{$Z_N(t)$} as $N\rightarrow+\infty$ \cl{(which yields a sort of ``law of large numbers'' for $Z_N$)} and the second step consists in specifying what happens around the mean \cl{("Central Limit Theorem")}. Heuristically, this \cl{can be} obtained \cl{ either by} expanding in Taylor series the generator ${\cal A}_N$ of $Z_N$ \cl{or by studying the paths  $(Z_N(t))$ by means of expression (\ref{PoissonN})}. 
We  sketch below the two perspectives for the behavior both at the mean and around it.\\
First, to  study $Z_N(t)$ at its mean, let us define the \cl{ the two functions $b_N $ and} $ b : [0,1]^p \rightarrow
{\mathbb R}^p$ by
\begin{equation}\label{defb}
 b_N(y)= \sum_{l \in E^-}\frac{l}{N} \alpha_l(Ny) \;\; \mbox{ and  }\;  b(y) = \sum_{l\in E^-}l \beta_l(y). 
\end{equation}
\cl{Under (H1), $b_N(.)$ converges  to $b(.)$ uniformly on $[0,1]^p$.}\\
Using definitions
(\ref{barZN}) and (\ref{defb}), the generator
${\cal A}_N$ of $Z_N$ writes, for $y \in E_N$,  \cl{ $f(.) \in C^1({\mathbb R}^p)$ with gradient  $\nabla f(y)$ and $l=\trans{(l_1,\dots,\ev{l_p})}$},
$${\cal A}_Nf(y)=   \sum_{l \in E^-} 
\alpha_l(N y)\biggl(f(y+\frac{l}{N})-f(y)-
 \frac{1}{N}\sum_{i=1,\dots,p} l_i \frac{\partial f}{\partial y_i}(y)\biggr)
+ b_N(y).\nabla f(y).$$
 \cl{Under (H1)}, the first term of $A_Nf(y)$ goes to 0, and the second term converges using (\ref{defb}) to $b(y). \nabla f(y)$. Therefore, \cl{as $N \rightarrow \infty$, $(Z_N(t))$ converges in distribution} to the process with generator ${\cal A}f(y)= b(y).\nabla f(y)$. 
\cl{The function $b$ inherits the regularity properties of $\beta_l$, so} $b$ is Lipschitz  by (H2) and the ODE (\ref{ODE}) has a unique well defined regular solution $x(t)$ satisfying
\begin{equation}\label{ODE}
x(t)= x_0 +\int_0^t b(x(u))du.
\end{equation}\\
\ev{Besides, }\cl{a  stronger result holds for $ (Z_N(.))$ (i.e. "law of large numbers").\\ 
If $Z_N(0) \tend{N}{+\infty}x_0$, then, under (H1)-(H2),}\\ 
\begin{equation}\label{CVU}
	\forall t \geq 0, \lim_{N\rightarrow \infty} \sup_{u\leq t} \parallel
Z_N(u)-x(u) \parallel=0\;\mbox{  a.s.}
\end{equation}\\

\noindent
Second, to specify the asymptotic behavior of the process
\cl{$(Z_N(t))$} around its deterministic limit $x(t)$, we have to pursue our approach for the mean by either expanding further $A_N$, or by setting a ``Central Limit Theorem'' for
\cl{$(Z_N(t))$}. The two approaches lead to different approximations. Indeed, the first one leads to a diffusion process $X_N (t)$, whereas the second approach \cl{ developed in Chapter 11, Section 2 of  \cite{eth05} consists in studying the process $ \sqrt N (Z_N(t)-x(t))$ based on the expression (\ref{PoissonN}) of $Z_N(.)$ in the specific case of transition rates $\alpha_l(.)$ such that 
\begin{equation}\label{HypEK}
\forall  l \in E^-,\;\forall y \in [0,1]^p, \; \forall N,\; \frac{1}{N} \alpha_l([Ny])=\beta_l(y).
\end{equation}}
\cl{Let us define  the two $p\times p$ positive symmetric matrices $\Sigma_N$ and $\Sigma$ from the $(\alpha_l)_{l\in E^-}$ and 
 $(\beta_l)_{l\in E^-}$ families. 
\begin{equation}\label{def:Sigma}
  \Sigma_N (y)=  \frac{1}{N}\sum _{l \in E^-} \alpha_l(y) l\; \trans{l}\; \; \mbox{ and } \;\; \Sigma(y)=  \sum _{l \in E^-} \beta_l(y)l\; \trans{l}.
\end{equation} } 
  
\noindent
Expanding  in Taylor series 
${\cal A}_N f(y)$ yields\cl{,} using (\ref{defb}) and (\ref{def:Sigma}),  for $f \in C^2 ( {\mathbb R}^p,{\mathbb R}),$
\begin{align*}
 {\cal A_N}f(y) &=\sum_{l \in E^-} \alpha_l(Ny)\biggl( f(y+\frac{l}{N})-f(y)-
\frac{1}{N} \sum_{i=1}^{d} l_i \frac{\partial f}{\partial y_i}(y)- 
\frac{1}{2 N^2} \sum_{i,j=1}^{p} l_i l_j \frac{\partial^2 f}{\partial y_i
\partial y_j}(y)
\biggr) \\
 &+ b(y).\nabla f(y) +  \frac{1}{2N}
\sum_{i,j=1}^{p} \Sigma_{ij}(y)
\;
\frac{\partial^2 f}{\partial y_i \partial y_j}(y) \\ 
&+ (b_N(y)-b(y)).\nabla f(y) +  
\frac{1}{2N}\biggl(
\sum_{i,j=1}^{p} (\Sigma_{N}-\Sigma)_{ij}(y)
\;
\frac{\partial^2 f}{\partial y_i \partial y_j}(y)\biggr).
\end{align*}\\
Heuristically, the first term of $ {\cal A}_Nf(y)$ is of $O(1/N^2)$, the second term  corresponds to the ODE; the second and third  terms correspond to the generator of \cl{a}  $p$-dimensional diffusion process \cl{$(X_{N}(t))$} with drift function $b(y)$ defined in \eqref{defb} and diffusion matrix $\Sigma(y)$ defined in \eqref{def:Sigma}. \cl{ Under (H1)}, the last term of ${\cal A}_Nf(y)$ is of order 
$o(\frac{1}{N})$. 
\cl{An additional  assumption on the $\alpha_l$ is required} to ensure that the remaining term 
$(b_N(y)-b(y)).\nabla f(y)$ is also $o(\frac{1}{N})$. 
\cl{Let us define more precisely the diffusion $X_N$.} Let $(B(t)_{t\geq 0})$  be a $p$-dimensional standard Brownian motion on a probability space ${\mathbb P}=(\Omega,({\cal F}_{t})_{t \geq 0}, P)$. Assume that   $Z_N(0)$ is  
${\cal F}_0$-measurable, then \cl{$X_{N}$ will be the solution of the
stochastic differential equation,}
\begin{equation}\label{XN}
dX_N (t)= b(X_N (t))\; dt + \frac{1}{\sqrt N} \;\sigma(X_N (t)) dB\cl{(t)};\;\; 
X_N (0)= Z_N(0), 
\end{equation}
\cl{where  $\sigma(y)$ is a $p \times p$ matrix satisfying 
$\sigma(y)^t \sigma (y) =\Sigma (y).$}\\
This is a Markov process with generator $B_N$ such that, \cl{for $f\in C^2(\R^p)$},
$$ B_Nf(y)= \frac{1}{2N} \sum_{i,j=1}^{p} \Sigma_{ij}(y) \frac{\partial^2 f}{\partial y_i \partial y_j}(y) + b(y).\nabla f(y).$$ 
\cl{The two generators ${\cal A}_N$ and $B_N$ satisfy $\parallel A_Nf-B_Nf \parallel =o(1/N)$, \ev{which} suggests the approximation of  $(Z_N(.))$ by  $(X_N(.))$.} \\
\cl{ Let us now briefly recall the results of  \citet{eth05} (Theorem 1, Chapter 11, Section 3) concerning $Y_N(t)= \sqrt{N}(Z_N(t)-x(t))$}. For this, define $\Phi(t,u)$, the resolvent matrix of the ``linearized'' ODE satisfied by $x(.)$, where
$\nabla b (y)$ denotes the \cl{$p\times p$} matrix $(\frac{\partial b_i}{\partial y_j}(y)$:
\begin{equation}\label{def:phi}
		\frac{d\Phi}{dt}(t,u)=\nabla b(x(t))\Phi(t,u) \;\; ; \; \; \Phi(u,u)=I_p.
	\end{equation}
\cl{ Then, for transition rates satisfying (\ref{HypEK}), \ev{\citet{eth05}} use the Poisson decomposition  (\ref{PoissonN}) to get that $(Y_N(t))$ converges in distribution to a centered Gaussian process} $(G(t))$, 
with covariance matrix, 
\begin{equation}\label{G}
	Cov(G(t),G(r))=\integ{0}{t \wedge r}\Phi(t,u)\Sigma(x(u))\trans{\Phi(\cl{r},u)}du.
\end{equation}\\
 \cl{The rigorous proof that $(Y_N(t))$ converges in distribution to $(G(t))$ in the case where the $ (\alpha_l(.))_l$ just satisfy  (H1) \ev{and} (H2) is surprisingly difficult. \ev{Moreover, the  Poisson decomposition  (\ref{PoissonN})} can no longer be used for the time-dependent processes that we consider in  the next section.  \ev{Hence, we prove the diffusion approximation for non homogeneous systems} using convergence theorems for semimartingales under the assumption:\\}
\noindent
{\bf (H1)'}: $\forall l \in E^-,
\;\; 
\super{y\in[0,1]}\norm{\cl{\sqrt{N}}\left(\frac{1}{N}\alpha_l([Ny])
-\beta_l(y)\right)}\tend{N}{\infty}0$.\\
 Some details along with a generalization  are given in Appendix \ref{App:Generalization}. \cl{Note that condition (H1)' does not seem to ensure that $ \parallel {\cal A}_N- B_N\parallel  =o(1/N)$ (where $\sqrt{N}$ has to be replaced by $N$). But the proof given in the appendix, which is more precise, requires only  (H1)’.}\\
\cl{Let us now set the different links between these two limit processes.}
Although the ``Central Limit Theorem'' and the generator expansion approach result in two different limiting processes, \cl{the theory of random pertubations of dynamical systems (or the stochastic Taylor expansion} of the diffusion, \cl{\citet{fre84,aze82})} clarifies the link between the Gaussian process $(G(t))$ \cl{defined in  (\ref{G})} and the diffusion process
\cl{$(X_N (t))$ defined in (\ref{XN}). Indeed, setting $\epsilon= 1/\sqrt{N}$, the paths  $X_N(.)$ satisfy,
\begin{equation}\label{TS}
	X_N (t) =X_{\epsilon}(t)= z(t)+ \epsilon g(t)+ \epsilon R_{\epsilon}(t),
\end{equation}
where $sup_{t\leq T} \parallel \epsilon R_{\epsilon}(t)\parallel \rightarrow 0 $ in probability as $\epsilon \rightarrow 0$, and where $z(t), g(t)$ are defined \ev{as follows}.  The function $z(t)$ satisfies the ODE $\frac{dz}{dt}= b(z(t))dt\;;\; z(0)=x_0$.\\
Therefore this is precisely the previous solution $x(t)$. The process $g(t) $ satisfies the stochastic differential equation
$$ dg(t)= \nabla b(x(t))g(t)dt +\sigma(x(t)) dB(t)\;\;;\;\; g(0)=0.$$
This SDE can be solved explicitely, and its solution is the process
$$g(t)= \int_0^t\Phi(t,s)\sigma(x(s)) dB(s).$$
Hence, $g(t)$ is a centered  Gaussian process having the same covariance matrix as the Gaussian process $G$ defined in (\ref{G}).}\\
\cl{For statistical purposes, this result is very useful. Indeed, it} was the starting point of the results in \cite{guy12}.\\

\noindent
We can summarize the approximation of our epidemic diffusion model for statistical purposes in the following algorithm. From now on, $(\alpha_l)$ and the derived functions will depend on parameters ($\theta$). 
\begin{description}
\item[Step 1:] Write all the mechanistic transitions between states and their
respective intensities (functions $\alpha_l$).
\item[Step 2:] Compute associated functions $\beta_{l}$ derived from (H1).
\item[Step 3:] Compute functions $b(\theta,y)$ and $\Sigma(\theta,y)$ (defined in \eqref{defb} and \eqref{def:Sigma} respectively) from $\beta_l$.
\end{description}

\subsection{Building the diffusion approximation for the $SIR$ epidemic model}
\label{SIR_appli}

We consider the simple $SIR$ model defined in Section \ref{intro} through the bidimensional continuous-time Markov jump process $Z_N(t)$. Following the three-step algorithm introduced above, we build the associated $SIR$ diffusion process.

{\bf Step 1:} The process $Z_N(t)$ has the state space $\{0,\dots, N\}^2$ and the jumps
$l$ are $(-1,1)$ and $(0,-1)$. The transition rates are respectively\\
 $q_{(S,I),(S-1,I+1)} =  \lambda S\;\frac{I}{N}=\alpha_{(-1,1)}(S,I)$ and $q_{(S,I),(S,I-1)} = \gamma I =\alpha_{(0,-1)}(S,I).$

{\bf Step 2:} Let $y=(s,i) \in [0,1]^2$ \ev{and} the parameter $\theta=(\lambda,\gamma)$ . Then,\\
 $\frac{1}{N}\alpha_{(-1,1)}([Ny])=  \frac{1}{N} \frac{\lambda}{N} [Ns][Ni]  \tend{N}{+\infty} \beta_{(-1,1)}(s,i)= \lambda s i$ ;\\$
\frac{1}{N}\alpha_{(0,-1)} ([Ny])=\frac{1}{N}\gamma[Ni]\tend{N}{+\infty}\beta_{(0,-1)}(s,i)= \gamma i.$\\
(H1)-(H2) are satisfied.

{\bf Step 3:} Function  $b(\theta,y)$ defined in (\ref{defb}) is then\\
 $b((\lambda,\gamma),(s,i))= \lambda si \begin{pmatrix} -1\\1
\end{pmatrix}+\gamma i \begin{pmatrix} 0\\-1 \end{pmatrix}= \begin{pmatrix}
-\lambda si \\ \lambda si -\gamma i \end{pmatrix}.$\\ 
The diffusion matrix  $\Sigma(\theta,y)$ defined in (\ref{def:Sigma}) writes as\\
$\Sigma((\lambda,\gamma),(s,i))=\lambda si \begin{pmatrix} -1\\1 \end{pmatrix} \begin{pmatrix} -1 & 1 \end{pmatrix} +\gamma i \begin{pmatrix} 0\\-1 \end{pmatrix} \begin{pmatrix} 0& -1 \end{pmatrix} = \begin{pmatrix} \lambda si & -\lambda si \\ - \lambda si & \lambda si + \gamma i \end{pmatrix}.$\\
\cl{Assume that} $(\frac{S_0}{N},\frac{I_0}{N})
\tend{N}{+\infty}(s_0,i_0)=x_0$, \cl{and let $\sigma(\theta,y)$ be a square root of 
$\Sigma (\theta,y)$}\\
\cl{and $(B(t)= \trans({B_1(t),B_2(t)})$ denote a standard two-dimensional Brownian motion.
Then, setting  $\sigma(\theta,(s,i))= \begin{pmatrix}
\sqrt{\lambda si} & 0 \\ -\sqrt{\lambda si} & \sqrt{\gamma i} \end{pmatrix}$
yields that $ X_N (t)= \begin{pmatrix}S_N (t)\\ I_N (t) \end{pmatrix}$ satisfies
$X_{N}(0)=x_0$} and\\
$\left\{\begin{array}{rl}
dS_N (t) & = -\lambda S_N (t) I_N (t) dt+\frac{1}{\sqrt{N}}\sqrt{\lambda S_N (t) I_N (t) }dB_1(t)\\
dI_N (t) & = (\lambda S_N (t) I_N (t)-\gamma I_N (t)) dt -\frac{1}{\sqrt{N}}\left(\sqrt{\lambda S_N (t) I_N (t)}dB_1(t)-\sqrt{\gamma I_N (t) } dB_2(t)\right).
\end{array}\right.$

\subsection{The diffusion approximation for the non autonomous case: the $SIRS$ model with seasonal forcing}
\label{time_dep}

While the $SIR$ model is suited for studying a single outbreak, it is not appropriate for describing recurrent epidemics. In order to reproduce a series of outbreaks, we need to compensate the depletion of susceptibles by other mechanisms, such as demography (with birth and death rates equal to $\mu$ for a stable population of size $N$) and/or reinsertion of removed individuals into $S$ compartment (as a consequence of immunity waning, after, on average, $1/\delta$ time). This leads to the $SIRS$ model. We also add a new term to the transition $S\rightarrow I$ that writes now as $(S,I)\stackrel{\frac{\lambda(t)}{N}S(I+N \eta)}{\longrightarrow}(S-1,I+1)$. This modification is introduced in order to avoid extinction, more likely to occur when simulating recurrent epidemics based on Markov jump process. The new term can be interpreted as constant immigration flow at rate $\eta$ in the infected class.\\
The diffusion approximation of this model obtained according to the scheme introduced in Section \ref{generic} is:

{\bf Steps 1 \& 2:}  $(S,I)\stackrel{\frac{\lambda}{N}S(I+N \eta)}{\longrightarrow}(S-1,I+1)$ $\Rightarrow \beta_{(-1,1)}(s,i)=\lambda s (i+\eta)$, \\
$(S,I)\stackrel{(\gamma+\mu) I}{\longrightarrow}(S,I-1)$ $\Rightarrow \beta_{(0,-1)}(s,i)=(\gamma+\mu)i$, \\
$(S,I)\stackrel{\mu S}{\longrightarrow}(S-1,I)\; \Rightarrow \beta_{(-1,0)}(s,i)=\mu s $ and\\
$(S,I)\stackrel{\mu N+\delta(N-S-I)}{\longrightarrow}(S+1,I)\;\Rightarrow
\beta_{(1,0)}(s,i)=\mu+\delta(1-s-i).$\\

\begin{figure}[ht]
\includegraphics[width=0.99\textwidth]{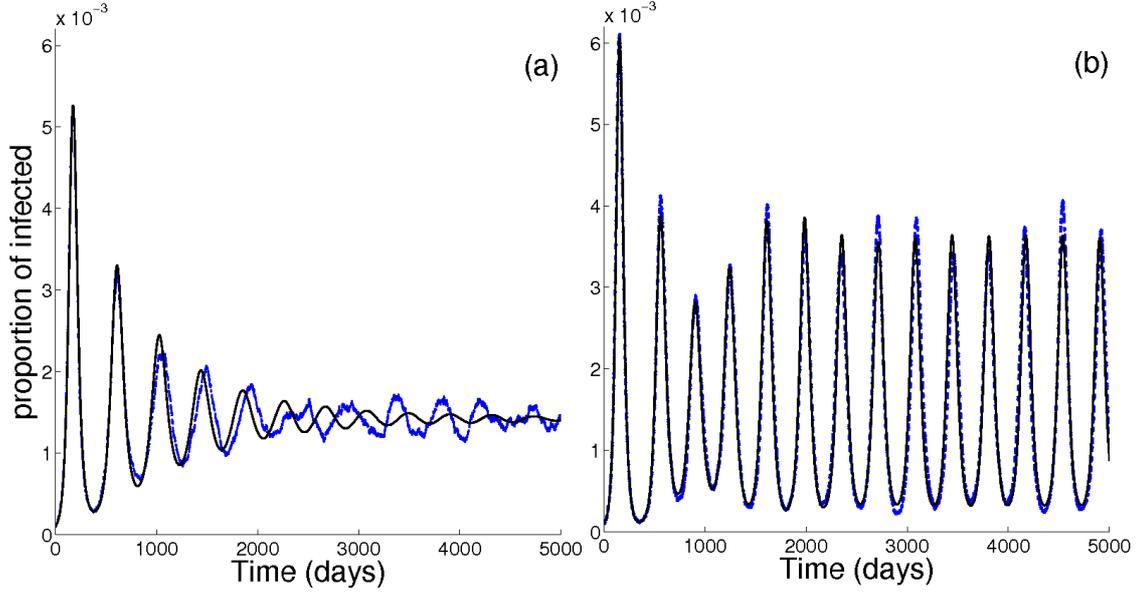}
	\caption{\label{fig:SIRS_ODES}Proportion of infected individuals over time for the diffusion approximation (blue) and the corresponding ODEs (black) of the $SIRS$ model with $N=10^7$, $T_{per}=365$, $\mu=1/(50 \times T_{per})$, $\eta=10^{-6}$, $(s_0,i_0)=(0.7,10^{-4})$ and $(\lambda_0,\gamma,\delta)=(0.5,1/3,1/(2\times 365))$, without seasonality, $\lambda_1=0$ (a) and with seasonality, $\lambda_1=0.02$ (b).}
\end{figure}

\noindent
We can notice that in the $SIRS$ population dynamics with multiple epidemic waves, the proportion of infected \ev{individuals} (the signal) is generally low ($\sim 10^{-3}$). Consequently, in order to obtain a satisfying ratio (greater than 1) between signal and noise, it is necessary to consider large populations ($N\sim 10^6$), since the noise has an order of $1/\sqrt{N}$.\\
Although able to describe more than one epidemic wave, it is well known that the $SIRS$ model leads to a function $b(y)$ and its associated ODE solution $(s(t),i(t))$, for which oscillations vanish (\cite{kee11}, Chapter 5) as $t\rightarrow \infty$ (and so does the diffusion, Figure \ref{fig:SIRS_ODES}).\\
\noindent
To overcome this problem, a natural assumption to be considered is that the transmission is seasonal. Mathematically, this leads to a time non homogeneous transmission rate, expressed under a periodic form \begin{equation}\label{lambdat}\lambda(t):=\lambda_0(1+\lambda_1
sin(2\pi t/T_{per}))\end{equation}
where $\lambda_0$ is the baseline transition rate,
$\lambda_1$ the intensity of the seasonal effect on transmission and $T_{per}$ the period of the seasonal trend (see
\citet{kee11}, Chapter 5).\\
\cl{ These transition rates define a time-inhomogeneous Markov jump process, for which 
the martingale problem is solved by means of an extended generator (\cite{eth05}, Chapter 4, Section 7). However,}  \ev{these authors} studied diffusion and Gaussian approximation\cl{s} of  Markov jump process\cl{es} only in the case of time homogeneous transition rates. 
\cl{Using that density dependent Markov processes can be viewed as semimartingales with specific characteristics, we studied the limit processes with another approach,
based on  convergence theorems for semimartigales (\cite{jac87}).}  
(see Appendix \ref{App:Generalization} for details).\\ 
Assumptions (H1), (H2), \cl{(H1)'} are modified as follows\cl{:}\\
{\bf (H1t)}: $\forall (l,t,y) \in E^- \times [0,T] \times  [0,1]^p,
\;\; 
\frac{1}{N} \alpha_l(t,[Ny]) \tend{N}{+\infty}\beta_l(t,y)$,\\ 
{\bf (H2t)}: $\forall l\in E^-, \; \beta_l (.,.)\in C^2([0,T]\times[0,1]^p).$\\
\noindent
{\bf (H1t)'}: $\forall l \in E^-,
\;\; 
\super{y\in[0,1],t\in[0,T]}\norm{\cl{\sqrt{N}}\left(\frac{1}{N}\alpha_l(t,[Ny])
-\beta_l(t,y)\right)}\tend{N}{\infty}0$.

\noindent
The new system obtained is still of dimension 2 (assuming a constant population
size) with four transitions for the corresponding Markov jump process. The procedure introduced in Section \ref{generic} can be generalized to time dependent models:

{\bf Step 1:}
$(S,I)\stackrel{\frac{\lambda(t)}{N}S(I+N\eta)}{\longrightarrow}(S-1,I+1)$,\\
$(S,I)\stackrel{\mu S}{\longrightarrow}(S-1,I)$,\\
$(S,I)\stackrel{(\gamma+\mu)I}{\longrightarrow}(S,I-1)$, and\\
$(S,I)\stackrel{\mu N+\delta(N-S-I)}{\longrightarrow}(S+1,I)$. 

{\bf Step 2:}
The rate of the first transition writes as\\
$\frac{1}{N}\alpha_{(-1,1)}(t,S,I)=\lambda(t)\frac{S}{N}\frac{I+N \eta}{N } \tend {N }{+\infty}\beta_{(-1,1)}(t,(s,i))=\lambda(t)s(i+\eta)$. \\
Since the time dependence satisfies $\alpha_l(t,k)=\lambda(t)\alpha'_l(k)$, and since only a space normalization is applied, the generic results from
Section \ref{generic} still hold by replacing functions $\beta_l(y)$ with
$\beta_l(t,y)$. The three other transitions are identical to those of the $SIRS$ model without seasonality.

{\bf Step 3:} Hence, for
$\theta=(\lambda_0,\lambda_1,\gamma,\delta,\eta,\mu)$, we obtain the drift term and diffusion matrix as:\\
$\begin{array}{ll}b(t,\theta,y)
&=\lambda(t) s(i+\eta) \begin{pmatrix}-1\\1\end{pmatrix}+(\gamma+\mu) i \begin{pmatrix}0\\-1\end{pmatrix}+(\mu+\delta(1-s-i))\begin{pmatrix} 1\\0\end{pmatrix} +\mu s \begin{pmatrix}-1\\0\end{pmatrix}\\
&=\begin{pmatrix}-\lambda(t) s(i+\eta) +\delta(1-s-i)+\mu(1-s)\\ \lambda(t) s(i+\eta)-(\gamma+\mu) i \end{pmatrix} \cl{= \begin{pmatrix} b_1(t,\theta,y)\\
b_2(t,\theta,y)\end{pmatrix},}
\end{array}$ 
$\begin{array}{lll}\Sigma(t,\theta,y)&=&\lambda(t) s(i+\eta)\begin{pmatrix} -1\\1 \end{pmatrix} \begin{pmatrix} -1 & 1 \end{pmatrix}
+(\gamma+\mu) i \begin{pmatrix} 0\\-1 \end{pmatrix} \begin{pmatrix} 0 & -1 \end{pmatrix}\\ & &
+ (\mu+\delta(1-s-i)) \begin{pmatrix} 1\\0 \end{pmatrix} \begin{pmatrix} 1 & 0 \end{pmatrix}
+\mu s  \begin{pmatrix} -1\\0 \end{pmatrix} \begin{pmatrix} -1 & 0 \end{pmatrix}\\&=&
\begin{pmatrix}\lambda(t) s(i+\eta)+\delta(1-s-i)+\mu(1+s)&-\lambda(t) s(i+\eta)\\-\lambda(t) s(i+\eta)&\lambda(t) s(i+\eta)+(\gamma+\mu) i\end{pmatrix}\end{array}$.\\

\noindent
\cl{The Cholesky} decomposition yields \cl{for $y=(s,i)$,} \\
\ev{
$\sigma(t,\theta,y)=\begin{pmatrix} \sigma_{1,1}(t,\theta,y) & 0\\ \sigma_{2,1}(t,\theta,y) & \sigma_{2,2}(t,\theta,y) \end{pmatrix}$ with\\
$\sigma_{1,1}(t,\theta,y)=\sqrt{\lambda(t) s(i+\eta)+\delta(1-s-i)+\mu(1+s)}$,\\
$\sigma_{2,1}(t,\theta,y)=-\sqrt {\frac{ \lambda(t) s(i+\eta)}{\lambda(t) s(i+\eta)+\delta(1-s-i)+\mu(1+s)}}$ and\\
$\sigma_{2,2}(t,\theta,y)=\sqrt{\frac{\lambda(t)(\gamma+\mu) si(i+\eta)+\lambda(t) s(i+\eta)(\delta(1-s-i)+\mu(1-s))+(\gamma+\mu)i(\delta(1-s i)+\mu(1+s))}{\lambda(t) s(i+\eta)+\delta(1-s-i)+\mu(1+s)}}$.\\}
\noindent
\ev{Finally, we get the diffusion  $X_N (t)= (S_N (t),I_N (t))$  starting from $ X_N(0)=(s_0,i_0)$,}
{\small\begin{equation}\label{model_SIRS}
\left\{\begin{array}{ccl}
\ev{dS_N (t)}&=&\ev{b_1(t,\theta;S_N (t),I_N (t))+\frac{1}{\sqrt{N}}\sigma_{1,1}(t,\theta;S_N (t),I_N (t))dB_1(t)}\\
\par \\
\ev{dI_N (t)}&=& \ev{b_2(t,\theta;S_N (t),I_N (t)) + \frac{1}{\sqrt{N}}\sigma_{2,1}(t,\theta;S_N (t),I_N (t))dB_1(t) + \frac{1}{\sqrt{N}}\sigma_{2,2}(t,\theta;S_N (t),I_N (t))dB_2(t).}
\end{array}\right.
\end{equation}}


\section{Minimum contrast estimators for diffusion processes}
\label{infer}

\cl{ The statistical inference for continuously observed diffusion processes on a finite interval is based on the likelihood of the diffusion and obtained using the Girsanov formula (see e.g. \cite{lip01} for the asymptotics $T\rightarrow \infty$ and \cite{kut84} in the asymptotics of small diffusion coefficient $\epsilon \rightarrow 0$). Discretely observed diffusion processes are discrete time Markov processes and thus their likelihood depends on the transition densities of the diffusion $p_{\theta}(t_{k-1},t_k ;x,dy)= 
\P_{\theta}(X(t_k) \in [y,y+dy]/ X(t_{k-1})=x)$. Since the dependence with respect to
the parameters $\theta$ of these transition densities is not explicit, the likelihood is untractable and other approaches have been proposed. This situation often occurs for stochastic processes, and other processes than the likelihood can be used to estimate parameters. These processes can be good approximations of the likelihood or can be completely different. For independent random variables, the estimators obtained with such approaches are called $M$-estimators (see \citet{vaa00}). For stochastic processes, the processes used instead of the likelihood are often called Contrast processes with associated Minimum contrast estimators. They have to satisfy a series of conditions to lead to good estimators. Contrary to the i.i.d. parametric set-up, there is no well recognized terminology to name these estimators, and we have adopted here the terminology of contrast processes and \ev{minimum contrast} estimators.}\\
In a previous work \citep{guy12} we developed a parametric inference approach for
discretely observed multidimensional diffusions with small diffusion
coefficient $\epsilon=1/\sqrt{N}$ (for $N$ large).  The diffusion is observed on interval $[0,T]$ at times $t_k=k\Delta$, for $k=1,..,n$ ($T=n\Delta$). We provided minimum
contrast estimators with good properties: consistent and asymptotically normal
for both drift and diffusion parameters for small sampling interval and for
drift parameters in the case of fixed sampling interval. Let us stress that, for general
diffusions with small diffusion coefficient observed on a fixed time interval, two different asymptotics can be considered. The first one corresponds to the small diffusion asymptotics ($\epsilon=1/\sqrt{N}\rightarrow 0$ $\Leftrightarrow$ $N\rightarrow +\infty$) and the second one corresponds to the sampling interval going to zero ($\Delta=\Delta_n\rightarrow 0$ $\Leftrightarrow$ $n\rightarrow +\infty)$. When the two asymptotics occur simultaneously, the rates of convergence of parameters in the drift and diffusion coefficient differ: drift parameters at rate $\epsilon^{-1}$ and diffusion parameters at rate $\sqrt{n}$. Consequently, for small sampling interval and for diffusion approximations where the same parameter $\theta$
is present in the drift and diffusion coefficients simultaneously, we can choose the most efficient rate to estimate this parameter. Here, we introduce a new variant of the contrast of \cite{guy12} (Section 3.3.1). \ev{This contrast is developed for the special case where the parameters of drift and diffusion terms are identical} \cl{ in} \ev{the asymptotics $N\rightarrow +\infty$ and for $n$ fixed. All these characteristics fit well the epidemic framework (e.g. large population size and limited number of observations). The new constrast} improves the non asymptotic accuracy of related estimators while preserving their asymptotic properties.

\subsection{The main lines of the inference method}
\cl{From now on, we assume that the parameter set $\Theta $ is a compact subset of $\R^m$, and that the true value of the parameter $\theta_0$ belongs to $\mathring{\Theta}$.}\\
As stated in previous sections, only the computation of functions $b$ and $\Sigma$ is required to build the approximation diffusion \eqref{XN} of the Markov jump process. These two functions allow building a
family of contrast processes for discrete observations \cl{at times} $t_k, k\in\{0,..,n\}$.\\
Using \eqref{defb},\eqref{ODE},\eqref{def:phi}, leads to $x_{\theta}(t)$, $b(\theta,x(t))$, the resolvent matrix
$\Phi_{\theta}$, and the Gaussian process $g_\theta(t)$ as
the limit of $\sqrt N ( {Z}_N(t)-x_\theta(t))$. Then, we can state the
fundamental property of our contrast approach. The Gaussian process $g_\theta$ satisfies:
\begin{equation}\label{eqg2}
g_\theta(t_k)=\Phi_\theta(t_k,t_{k-1})g_\theta(t_{k-1})+\sqrt{\Delta}V_{k}
^\theta,
\end{equation} 
with $(V_k^\theta)_k$ a sequence of $n$ independent centered Gaussian vectors with
bounded covariance matrix, and $\Delta=T/n$ the sampling interval. The sequence
$(V_k^\theta)$ being independent, we can compute its likelihood and derive a
contrast process for the diffusion. \\
For this, let us define the function $A_k(\theta,(X_{t_k})_{k\in\{0,..,n\}})=A_k(\theta)$ for the diffusion $(X_N (t))$ at time points $(t_k)_{k\in \{0,..,n\}}$,
\begin{equation}
 A_k(\theta)=X_{t_k}-x_\theta(t_k)-\Phi_\theta(t_k,t_{k-1})\left[X_{t_{k-1}} -x_\theta(t_{k-1})\right].
\end{equation}
Let us also introduce the matrix $S_k^\theta$, corresponding to the covariance matrix of the family $(V_k^\theta)$ as\\
$S_k^\theta=\frac{1}{\Delta}\integ{t_{k-1}}{t_k}\Phi_\theta(t_k,u)\Sigma(\theta,x_\theta(u))\trans{\Phi}_\theta(t_k,u)du$.\\
This leads to the construction of the contrast process $U_{N}$ and the associated estimator $\hat{\theta}_{N}$:
\begin{equation}\label{modif_cont}
\begin{array}{ll}
U_{N}(\theta,(X_{t_k})_{k\in\{0,..,n\}})=\som{k=1}{n} \left[\frac{1}{N}log\left(det\left(S_k^\theta\right)\right)+\frac{1}{\Delta} \trans{A_k(\theta) } \left(S_k^\theta\right)^{-1}A_k(\theta)\right],\\
\hat{\theta}_{N}=\underset{\theta \in\Theta}{argmin}\;U_{N}(\theta,(X_{t_k})_{k\in\{0,..,n\}}).
\end{array}
\end{equation}
In this contrast process, $A_k(\theta)$ can be interpreted as an error function
between observations and the deterministic trajectory associated to the
parameter $\theta$ at time $t_k$, incorporating the propagation of the error at
time $t_{k-1}$, and $S_k^\theta$ as a corrective weight matrix.
\noindent
The contrast \eqref{modif_cont} is a modified version of a contrast proposed
in our previous work (Section 3.3.1 of \cite{guy12}). The main improvement
is provided by the additional term
$\frac{1}{N}log\left(det\left(S_k^\theta\right)\right)$ in \eqref{modif_cont} 
which corrects a non asymptotic bias of $\hat{\theta}_{N}$ (noticed in
simulations presented in \cite{guy12}), while preserving its asymptotic
properties as $N\rightarrow \infty$ and $n$ fixed. Since $\som{k=1}{n}
\frac{1}{N}log\left(det\left(S_k^\theta\right)\right)$ is a finite sum of
bounded terms, it will tend to $0$ as  $N\rightarrow \infty$.  According
to Proposition 3.2 in \cite{guy12}, $\hat{\theta}_{N}$ is consistent and
asymptotically normal:\\
$\sqrt{N}\left(\hat{\theta}_{N}-\theta_0\right)\overset{\mathcal{L}}{\longrightarrow}\mathcal{N}_{\cl{m}}\left(0,I^{-1}(n,\theta_0)\right)$ where\\
$I(n,\theta_0)=\left(\som{k=1}{n}D_{k,i}\left(S_k^{\theta_0}\right)^{-1}\trans{D} _ { k , j } \right)_{\cl{1\leq i,j \leq m}}$\\ with   
$D_{k,i}=-\deriv{x_\theta(t_{k})}{\theta_i}(\theta_0)+\Phi_{\theta}(t_{k},t_{k-1} ) \deriv{x_\theta(t_{k-1})}{\theta_i}(\theta_0)$.\\
\noindent
It is important to point out that the above results are still valid for any
number of observations $n$. As $n$ increases, the asymptotic information
$I(n,\theta_0)$ increases (and consequently the width of confidence intervals
decreases) towards the efficient bound corresponding to the continuous observation of the
diffusion on $[0,T]$ for parameters in the drift functions:\\
$I(n,\theta_0)\tend{n}{\infty} I_b(\theta_0) ={\small \left(\frac{1}{T}\integ{0}{T}\deriv{b(\theta, x_\theta(t))}{\theta_i}(\theta_0)\Sigma^{-1}(\theta_0,x_{\theta_0}(t))\trans{ \deriv{b(\theta,x_\theta(t))}{\theta_j}(\theta_0)}dt\right)_{\cl{1\leq i,j\leq m}}}$.\\
\begin{rem}
For irregular sampling interval, $\hat{\theta}_{N}$ will still keep its properties
(see Appendix \ref{App:Non-autonomous} for more details). This aspect has
practical implications since it can be used in various observed designs of
epidemics: for instance, many data points could be recorded in the early phase of
the epidemic and much less in the second phase.
\end{rem}

\subsection{Case of time dependence}
As stated in Section \ref{time_dep}, the diffusion approximation holds for time non homogeneous Markov jump processes. This leads to  drift and diffusion
functions  $b$ and $\Sigma$ which are time dependent. Although the results obtained in \cite{guy12} were proved only for autonomous diffusions $dX_N (t)=b(\theta,X_N (t))dt+\frac{1}{\sqrt{N}}\sigma(\theta,X_N (t))dB_t,x_0\in \R^p$, they can extend to time dependent diffusion processes. Previous quantities need to be modified by replacing each occurrence of $b(\theta,x_\theta(t))$, $\Sigma(\theta,x_\theta(t))$ and $\sigma(\theta,x_\theta(t))$ by $b(t,\theta,x_\theta(t))$, $\Sigma(t,\theta,x_\theta(t))$ and $\sigma(t,\theta,x_\theta(t))$. The estimates of $SIRS$ model parameters are obtained using this new framework (Section \ref{simu:SIRS}).  Additional technical details are provided in Appendix \ref{App:Non-autonomous}.

\section{Simulation study}
\label{simul}
The inference method proposed in this study is assessed on simulated data.  Two different epidemic models, the $SIR$ and the $SIRS$ with time-dependent transmission rate and demography, described in Sections \ref{SIR_appli} and \ref{time_dep} respectively, are considered. Simulations are based on the algorithm of \citet{gil77} for the $SIR$ model and on the $\tau$-leap method (\cite{cao05}), more efficient for large populations, for the $SIRS$ model. The accuracy of our minimum contrast estimators is investigated with respect to the population size $N$, the number of observations $n$, the parameter values and the model generating the data (Markov jump process and diffusion process). \ev{Only non extinct trajectories are considered for inference.} An
ad-hoc criterion (final epidemic size larger than $5\% $ of the number of initial susceptibles) is chosen to define non extinction. For each set of parameter values, point contrast estimates ($CE$), theoretical confidence intervals ($CI_{th}$) and empirical ones ($CI_{emp}$, built on 1000 runs) are provided. Moreover, the intrinsic limits of the method are investigated by
comparing $CI_{th}$ for different values of $n$, other parameters being fixed, with the theoretical variance co-variance matrix when $n \rightarrow \infty$. \\
The first finding is that no relevant bias can be imputed to the model underlying the simulations, when comparing $CE$s on data simulated under Markov jump and diffusion processes. Therefore, all the subsequent investigations were performed based on simulated trajectories with the Markov jump process. \ev{To facilitate the visual comparison of theoretical ellipsoids, they are all centered on the true parameter values for each scenario.} Additionally, as $CI_{emp}$ are very tight around point estimators, they are not represented on figures.

\subsection{The $SIR$ model}
From now on, we consider the parameters of interest for epidemics: the basic reproduction number, $R_0=\frac{\lambda}{\gamma}$, which represents the average
number of secondary cases generated by one infectious in a completely
susceptible population, and the average infectious duration, $d=\frac{1}{\gamma}$. The performances of our $CE$s were evaluated for several parameter values. For
each combination of parameters, the analytical maximum likelihood estimator ($MLE$), calculated from the observation of all the jumps of the Markov process, was taken as reference.

\begin{table}[ht]
\begin{tabular}{|c|l|l|}
    \hline
    Parameter & Description & Values \tabularnewline
    \hline
    $R_0$ & basic reproduction number & 1.5, 3  \tabularnewline
    \hline
$d$ & infectious period & 3, 7 days  \tabularnewline
\hline
$T^{(1)}$ & final time of observation & 20, 40, 45, 100  days\tabularnewline
\hline
$N$ & population size & 400, 1000, 10000  \tabularnewline
    \hline
$n$ & number of observations & 5, 10, 20, 40, 45, 100  \tabularnewline
    \hline
 \end{tabular}
\caption{\label{range_param} Range of parameters for the $SIR$ model defined in Section \ref{SIR_appli}. $^{(1)}$: $T$ is chosen as the time point where the corresponding deterministic trajectory passes below the threshold of $1/100$.}
\end{table}

\begin{figure}[ht] 
\includegraphics[width=0.9\textwidth]{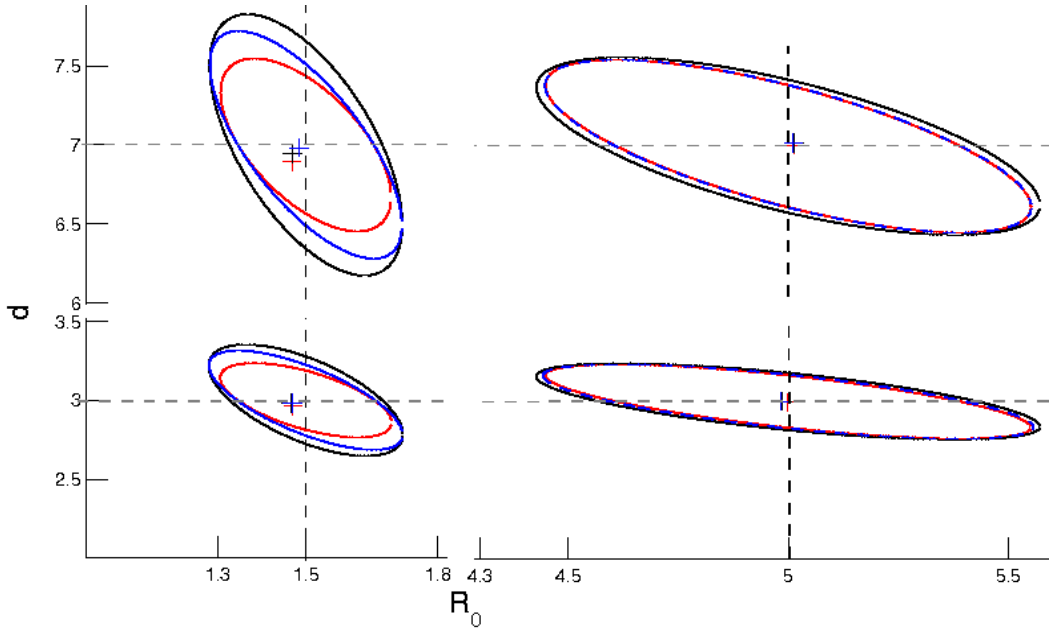}
\caption{\label{fig:1000_1}\ev{Point estimators (+) built as averages over 1000 independent simulated trajectories and their associated theoretical confidence ellipsoids  for the $SIR$ model: $MLE$ with complete observations (red), $CE$ for 1 obs/day, $n=40$ (blue)} and $CE$ for $n=10$ (black). Four scenarios are illustrated: $(R_0,d,T)=\{(1.5,3,40);(1.5,7,100);(5,3,20);(5,7,45)\}$, with $N=1000$. \ev{True parameter values are located at the intersection of horizontal and vertical dotted lines.}}
\end{figure}
\begin{figure}[ht]
\includegraphics[width=0.9\textwidth]{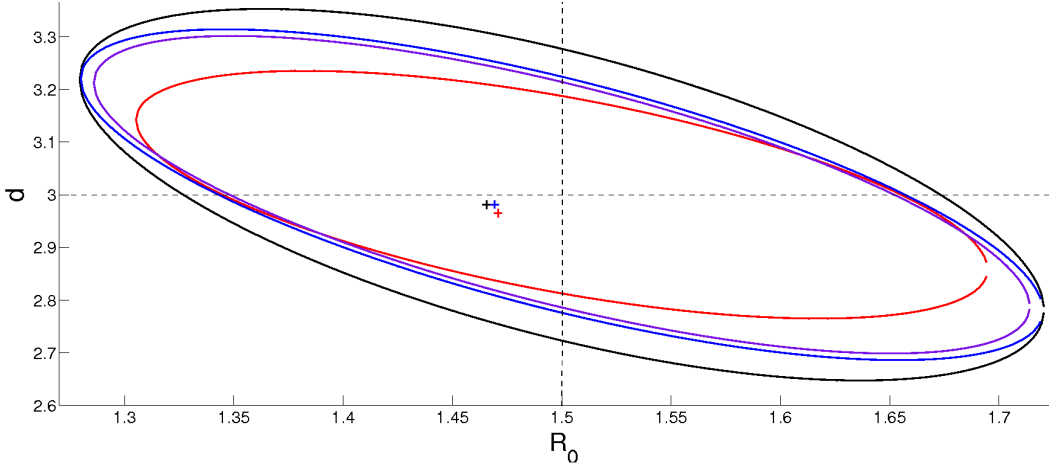}
\caption{\label{zoom_SIR}Zoom of Figure \ref{fig:1000_1} for $(R_0,d,T)=(1.5,3,40)$ \ev{with an additional theoretical confidence ellipsoid corresponding to $n=2000$ (purple).}}
\end{figure}

\textcolor{white}{a}\\
As a general remark, we can say that the magnitude of the stochasticity of the sample path of $I_N (t)$ depends on the value of $R_0$: for small $R_0$ the proportion of infected individuals in the population is smaller and so is the ratio signal over noise.\\
Figure \ref{fig:1000_1} illustrates the accuracy of the $CE$s for a moderate
population size $N=1000$ and from trajectories with weak ($R_0=5$) and strong
($R_0=1.5$) stochasticity. First, we can notice that there is a non negligible
correlation between parameters $R_0$ and $d$ (ellipsoids are deviated with respect to the $Ox$ and $Oy$ axes), increasing with $d$ and decreasing with $R_0$, even for the $MLE$. Second, the shape of confidence ellipsoids (and consequently 
the projection on $Ox$  and $Oy$ axes with the largest $CI_{th}$ among $R_0$ and $d$) depends on parameter values: e.g. $CI_{th}$ is larger for $R_0$ than for $d$ when $R_0=5$,
whereas the opposite occurs for $R_0=1.5$. Third, for $R_0=5$ (i.e. for
trajectories with weak stochasticity), all the $CI_{th}$ are very close (especially those of $MLE$ and $CE$ for 1 obs/day), suggesting that there is no loss in estimation accuracy
as expected for smooth trajectories, even when not all jumps are observed. This
does not stand for $R_0=1.5$ when trajectories are very noisy (Figure
\ref{zoom_SIR}): the shape of ellipsoids and their relative positions vary with
$n$. More specifically, for these trajectories, a large number of discretized observations ($n=2000$ which corresponds to the maximum number of possible jumps for $N=1000$ $SIR$ dynamic with two types of transitions) does not compensate the
loss of information
compared to the case where all dates of jumps are observed. Finally, point
values for $MLE$ and $CE$ calculated for different $n$ are very similar, which
recommends the use of our $CE$s when only a small number of observations are
available.\\

\begin{figure}[ht]
\includegraphics[width=0.9\textwidth]{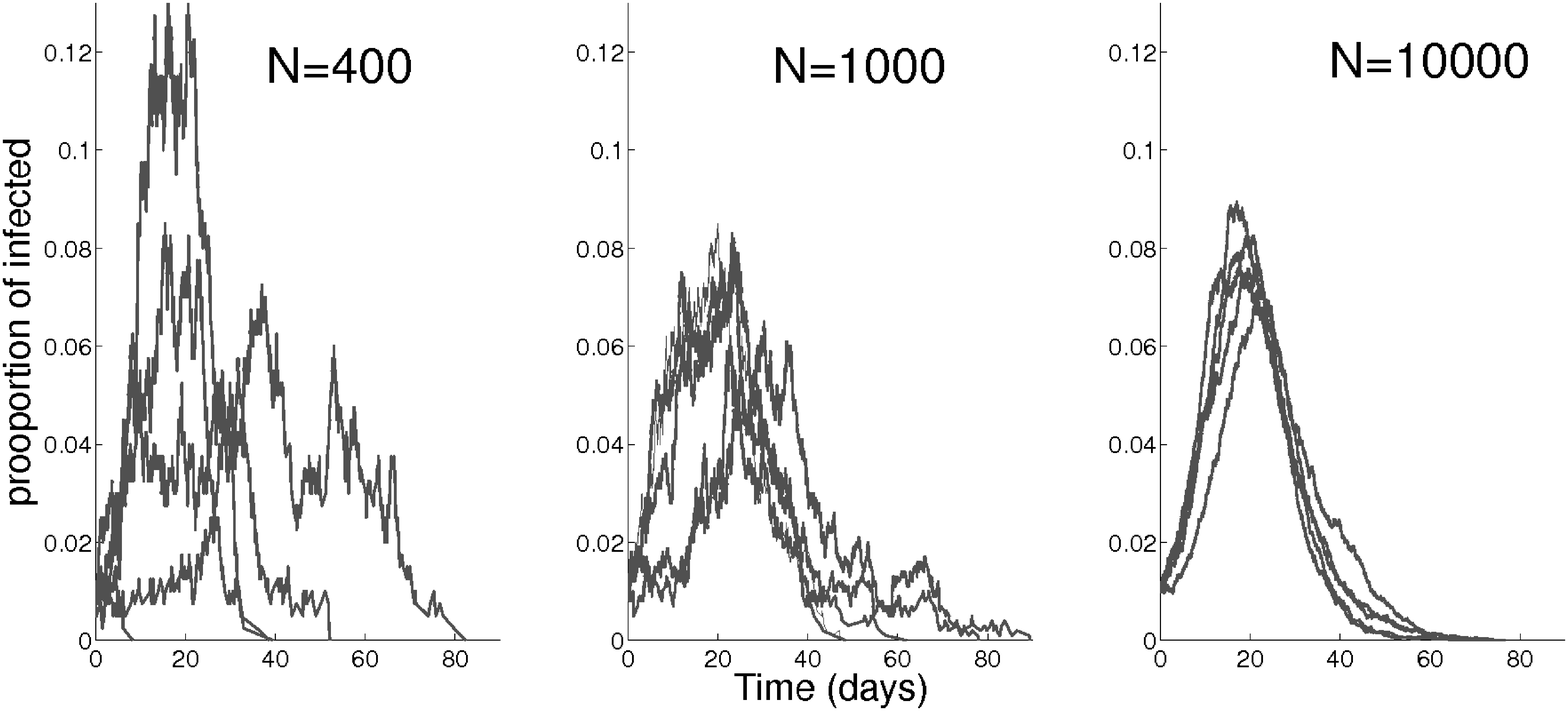}\\
\includegraphics[width=0.9\textwidth]{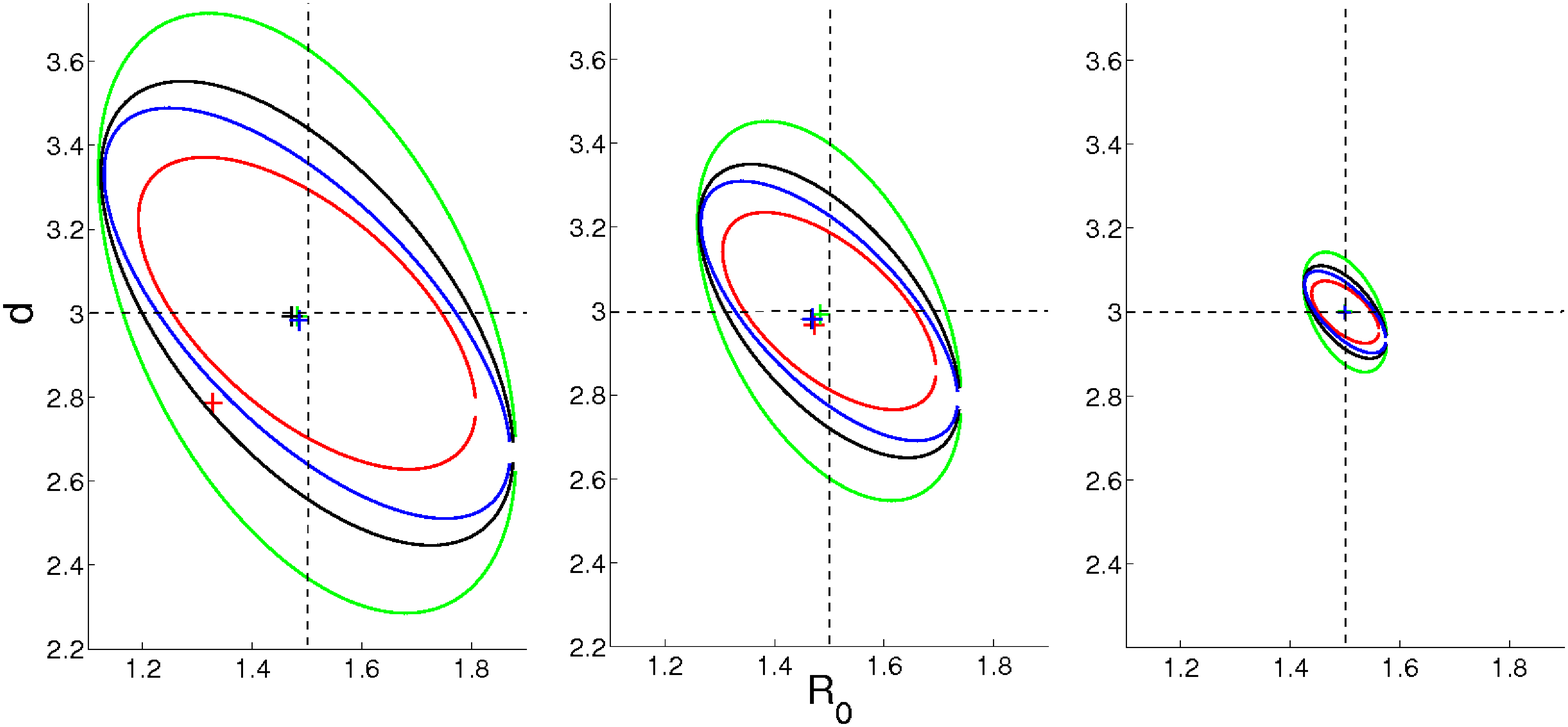}
\caption{\label{fig:400-10000} \ev{Several simulated trajectories of the proportion of infected individuals over time using the $SIR$ Markov jump process (top panels). Average point estimators (+) over 1000 independent simulated trajectories (same model) and their associated theoretical confidence ellipsoids (bottom panels)}: $MLE$ with complete observations (red), $CE$ for 1 obs/day, $n=40$ (blue), $CE$ for $n=10$ (black) and $CE$ for $n=5$ (green) for $(R_0,d)=(1.5,3)$ and $N=\{400,1000,10000\}$ (from left to right). \ev{True parameter values are located at the intersection of horizontal and vertical dotted lines.}}
\end{figure}

\begin{figure}[ht]
\includegraphics[width=0.9\textwidth]{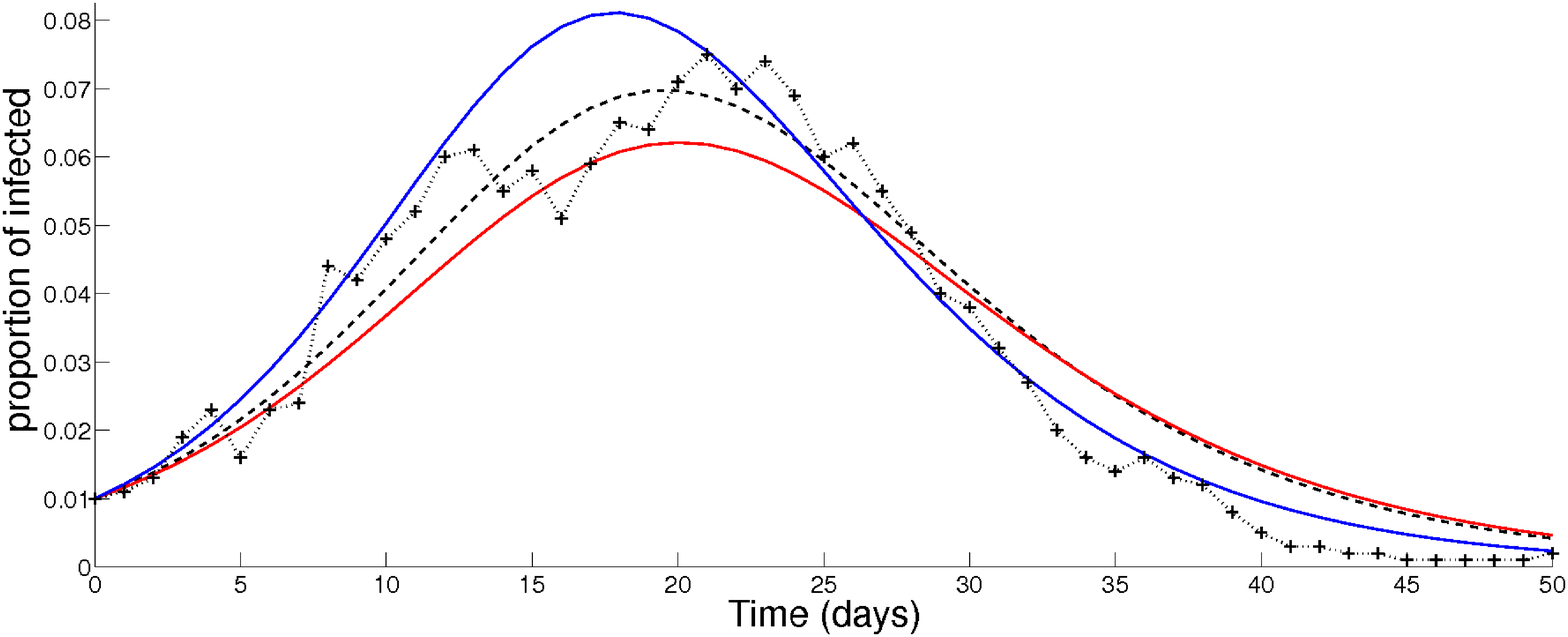}\
\caption{\label{fig:traj-estim} \ev{One simulated trajectory of the $SIR$ Markov jump process representing the simulated data (black line with crosses). Associated deterministic trajectories obtained with: the true parameter values, $(R_0,d)=(1.5,3)$ (dotted black line),} the $MLE$ with complete observations ($1.46$, $2.96$) (red) and the $CE$ ($1.56$, $2.91$) (blue).}
\end{figure}
\textcolor{white}{a}\\
Figure \ref{fig:400-10000} shows that the width of confidence intervals, when projecting on $Ox$ and $Oy$ axes, decreases with respect to $\sqrt{N}$, whereas the correlation is not impacted by $N$. For a given value of population size $N$, confidence ellipsoids are relatively close except for the case of very few observations (e.g. $n=5$). For the case $N=400$, the $MLE$ is biased, mostly due to the strong variability in the trajectories. An explanation of this behavior lies in the fact that the $MLE$ is optimal when data represent a "typical" realization
of the Markov process, but could exhibit a bias when observations are far from the mean. This does not seem to occur  when using our $CE$s. Although our method was introduced for large populations, it proves to be quite robust w.r.t. highly variable sample paths (obtained either for small $N$ or small $R_0$).\\
We can see on Figure \ref{fig:traj-estim} that even if on a large number of trajectories the asymptotic properties of $MLE$ and $CE$s are very similar, for a particular trajectory, the estimation accuracy may be different. Beyond the intrinsic variability of point estimates, this can also be viewed as a consequence of the form of functions $A_k(\theta)$. Indeed, they are more sensitive to variation in slope of the difference between deterministic trajectory and data than the classical least square distance.
\begin{rem}
The robustness of estimations to the misspecification of $N$ is an interesting point since the population size can be inaccurately known in practice. Assume that the true size of the population is $N$ and that the wrong value $N'$ has been used instead. Starting from the numbers of susceptible and infected individuals, we normalize by $1/N'$ these quantities and then build the estimators. For the $SIR$ model,  we still obtain the right estimator for $d$, while $R_0$ is no longer consistently estimated. Our procedure estimates instead the quantity $R_0 N/N'$. This would also occur with estimation based on the ODE, while estimators based on all the jumps of $(S(t),I(t))$ would not exhibit this bias.
\end{rem}

\FloatBarrier
\subsection{The $SIRS$ model}
\label{simu:SIRS}
\begin{figure}[ht]
\includegraphics[width=0.99\textwidth]{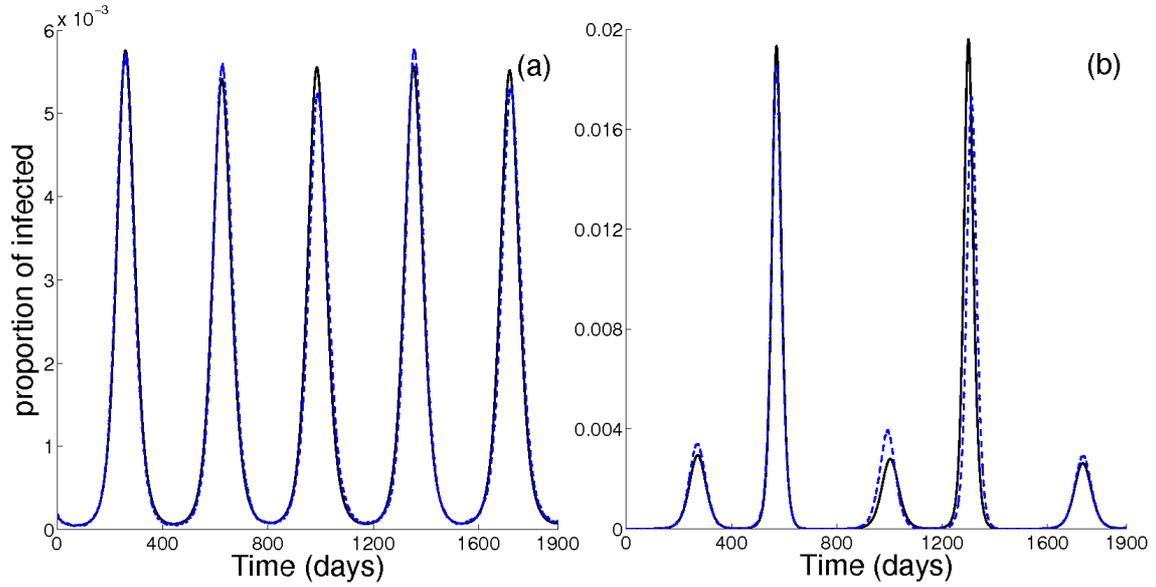}\\
\caption{\label{fig:SIRStraj} Deterministic (black) and Markov jump
process (blue) trajectories of the $SIRS$ model with demography and seasonality in transmission. Proportion of infected individuals over time for $N=10^7$, $(s_0,i_0)=(0.7;10^{-4})$ $\eta=10^{-6}$, $\mu=\frac{1}{50}$, $(R_0,d,\frac{1}{\delta T_{per}})=(1.5,3,2)$, and
(a) $\lambda_1=0.05$ and (b) $\lambda_1=0.15.$}
\end{figure}

\begin{figure}[ht]
\includegraphics[width=0.99\textwidth]{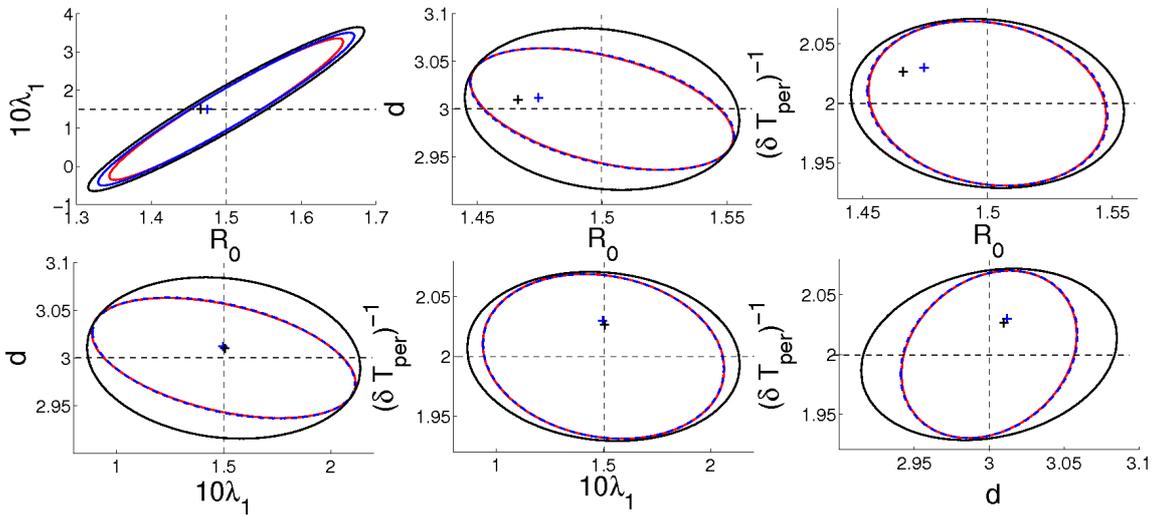}\\
\caption{\label{fig:SIRSafterbif} \ev{Point estimators (+) built as averages over 1000 independent simulated trajectories and their associated theoretical confidence ellipsoids  for the $SIRS$ model} with demography and seasonal forcing in transmission: $CE$ for 1 obs/day (blue) and for $n=1$ obs/week (black)
for $(R_0,d,\lambda_1,\delta)=(1.5,3,0.15,2)$, $T=20$ years and $N=10^7$. Asymptotic confidence ellipsoid ($n \rightarrow \infty$) is also represented (red). \ev{True parameter values are located at the intersection of horizontal and vertical dotted lines.}}
\end{figure}

For the $SIRS$ model describing recurrent outbreaks and defined in Section
\ref{time_dep}, four parameters were estimated: in addition to $R_0$ and $d$, $\lambda_1$ and $\delta$ were assessed (the latter ones were replaced in estimations, for numerical reasons, by $10\times \lambda_1$ and $1/\delta T_{per}$). Demographic parameter $\mu$ was fixed to $1/50$ $\mbox{years}^{-1}$, a value usually considered in epidemic models, $T_{per}$ was taken equal to $365$ days
and $\eta=10^{-6}$, which corresponds to $10$ individuals in a population size of $N=10^7$ . The large value of $N$ considered allows a sufficient pool of susceptible and infected individuals at the end of each outbreak for the epidemic to restart in the next season. Our $CE$s were assessed on trajectories obtained for
$(R_0,d,\lambda_1,\delta)=\{(1.5,3,0.05,2),(1.5,3,0.15,2)\}$ and $T=20$ years. The two scenarios correspond to $\lambda_1$ respectively smaller and larger than the bifurcation point of the corresponding ODE system (see \cite{kee11}, Chapter 5.1, for more details on the bifurcation diagram of the $SIRS$ deterministic model). For numerical scenarios considered, the bifurcation value for $\lambda_1$ is around $0.07$. However the qualitative pattern of epidemic dynamics (from annual to multiannual epidemics) also depends on the remaining parameter values (in particular, $\eta$ seems to have an important impact). As depicted in Figure \ref{fig:SIRStraj}, for $\lambda_1=0.05$ the proportion of infectives exhibits oscillations which are roughly annual and of constant amplitude, whereas for $\lambda_1=0.15$ dynamics are biennial. Numerically, the scenarios considered have the characteristics of influenza seasonal outbreaks. According to results in Figure \ref{fig:SIRSafterbif}, illustrating different projections of the four-dimensional theoretical confidence ellipsoid, almost no correlation is noticed between estimators, except for $R_0$ and $\lambda_1$ after bifurcation. Moreover, the accuracy of estimation is relatively high, regardless to the parameter. Interestingly, disposing of 1 obs/day (which can be considered as a practical limit of data availability) leads to an accuracy almost identical to the one corresponding to a complete observation of the epidemic process (blue and red ellipsoids respectively in
Figure \ref{fig:SIRSafterbif}). Estimations based on 1 obs/week provide less but still reasonably accurate estimations.

\noindent
The width of $CI_{th}$ are similar for the two scenarios tested except for the parameter $\lambda_1$ (Figure \ref{fig:SIRSafterbif} for $\lambda_1=0.15$ and Figure S1 for $\lambda_1=0.05$). However, correlation between parameters (mainly $(R_0,\lambda_1)$ and $(d,1/\delta T_{per})$) may vary according to the value of $\lambda_1$. The shape of the ellipsoid for $(d,1/\delta T_{per})$ is also slightly different between the values tested and an $SIRS$ model without seasonality (see electronic supplementary material, Figure \ref{fig:SIRSafterbif}, Figure S1 and Figure S2). This can be partly explained by qualitative differences in corresponding deterministic dynamics of infected individuals. Finally, here again, one observation per day leads to remarkably accurate estimations.

\FloatBarrier
\section{Conclusion}
\label{conc}

\ev{In this study we provided first a rigorous and easy to implement three-step method for elaborating the diffusion approximation of Markov jump processes describing epidemic spread. Then, we developed a contrast-based inference method for parameters of epidemic models represented by diffusion processes, when all the coordinates of the system are discretely observed. The elaboration of the diffusion approximation builded on results of \citet{eth05}, but using a different technique as their time changed Poisson representation. Namely, we showed that the jump process and the diffusion both converge towards the same Gaussian process. Since our generalization encompasses time inhomogeneous systems, it allows handling complex epidemic models, particularly those with time-dependent transition rates. Our estimators have good properties for asymptotics corresponding to realistic situations in epidemiology, especially for large populations and limited number of observations. We also showed that a correction term introduced in the contrast function theoretically improved the non asymptotic accuracy of the estimators while preserving their asymptotic properties. In addition, we proved that our estimators were well fitted for models including both time homogeneous and non homogeneous transition rates. Performances of our estimators and their robustness with respect to parameter values were investigated on simulated data. The estimation accuracy depends on the variablity of trajectories (impacted by the population size and the basic reproductive number). However, even for noisy dynamics, our estimators behave noticeably well for realistic scenarios: one observation per day allows obtaining an accuracy close to that of the estimator for complete data (continuous observation).}\\
These promising findings lay the foundations of an inference method for partially (in time and state space) observed epidemic data, a more realistic scenario. The main interest of our method, developed for time discrete observations (partial in time) of a continuous process is the fact that it is mostly an analytic approach, requiring only the classical optimization steps. It should be viewed as a complementary approach to the powerful current inference techniques for partially observed processes, which necessitate computer intensive simulations \ev{for data completion and numerous tuning parameters to be adjusted.  Indeed, our method can provide first estimations to initialize these algorithms, which, in turn, can tackle more sophisticated epidemic models.}

\section{Appendix}
\label{App}
\subsection{Time changed Poisson process representation of a Markov jump process}
\label{App:Poisson_repr}
First, the process satisfying (\ref{Poisson}) is obtained recursively as
follows. 
Let  $Z_0(t)\equiv Z(0)$ and set \\$Z_1(t)= Z(0)+ \sum_{l \in \Z^p} l\;
P_l\bigg(\int_0^t \alpha_l(Z_{0}(u))du\biggr).$
For $k >1$, define \\$Z_k(t)= Z(0)+ \sum_{l \in \Z^p} l\;
P_l(\int_0^t \alpha_l(Z_{k-1}(u))du).$
Then, if $\tau_k$ is the $k$th  jump of $Z_k(t)$,  \\$Z_k(t)= Z_{k-1}(t)$ for
$t< \tau_k$.  
The process $Z(t)= \underset{k\rightarrow \infty}{lim}\; Z_k(t)$ exists and
satisfies (\ref{Poisson}).\\
Second, a characterization of these random time changed processes is mainly
based on the property: given a positive measurable function $\eta: E \rightarrow
(0, +\infty)$ and a Markov process $Y(.)$ such that
$\int_0^{\infty}\frac{du}{\eta(Y(u))}= \infty $ a.s., one can define the random
time change $\tau(t)$ by $\int_0^{\tau(t)}\frac{du}{\eta(Y(u))} =t 
\Longleftrightarrow \dot{\tau}(t)= \eta(Y(\tau(t)).$ 
The process $R(t)$ defined as $ R(t) := Y(\tau(t))$ satisfies the equation
$R(t) =Y(\int_0^t \eta(R(u))du)$. Moreover, if $A$ is the generator of $Y(.)$,
the generator of $R(t)$ is equal to $\eta A$. 
Now, if $(Y(t))$ is the Poisson process $P_l(t)$ with rate $1$ (generator 
$Af(k)= f(k+1)-f(k)$), and  $\eta (.)= \alpha_l(.)$, the process 
$Z_l(t) = P_l( \tau_l(t))$ has generator $A_lf(k)= \alpha_l(k)(f(k+1)-f(k))$ and
satisfies, $Z_l(t) =P_l(\int_0^t \alpha_l(Z_l(s))ds)$.
This allows to prove that the solution of (\ref{Poisson}) has the generator $Af(k) =\sum_{l \in \Z^p} \alpha_l(k)(f(k+l)-f(k))=\alpha(k)\sum_{l \in \Z^p}
(f(k+l)-f(k))\frac{ \alpha_l(k)}{\alpha(k)}.$ We identify this generator as the one of $(Z(t))$ defined by \eqref{trans}.

\subsection{Diffusion approximation for non-homogeneous Markov jump processes}
\label{App:Generalization}
We extend the approximation results from \cite{eth05} to the time dependent case. Their approach consists in using a Poisson time changed representation of the Markov jump process, a Brownian motion time changed representation of the diffusion process, and to compare them with an appropriate theorem from \cite{kom76}. The extension of the proof of \cite{eth05} detailed in Appendix \ref{App:Poisson_repr} relies on the existence of \eqref{Poisson} for time dependent Markov processes. The main problem is that the natural characterization of the random time change stated in Appendix \ref{App:Poisson_repr} now writes
$\int_0^{\tau(t)}\frac{du}{\eta(\tau^{-1}(u),Y(u))} =t $, and the
time change becomes implicit. We rather use the general convergence results from \cite{jac87} to obtain
the diffusion approximation.\\
We consider the pure Jump Markov process $\cl{Z}(t)$ with state space $E=\{0,..,N\}^p$ and transitions rates $q_{x,x+l}(t)=\alpha_l(t,x)$. This process has for generator\\ $\mathcal{A}_tf(x)=\integ{\R^p}{} K_t(x,dy)\left(f(x+y)-f(x)\right)$ with the transition \cl{k}ernel\\ $K_t(x,dy)=\som{l\in E^-}{}\alpha_l(t,x)\delta_l(y)$ \cl{where $\delta_l$ is the Dirac measure at point $l$.} \\
Within the framework  developed by \cite{jac87}, it is a semimartingale \cl{with a random jump measure integrating $\parallel y\parallel$}. So its characteristics in the sense of Definition 2.6 in Chapter II are  
\cl{$(B,C,\nu_t)$, where\\
1. $B=(B_i(t))_{1 \leq i \leq p}$ is the predictable process, $B(t)=\int_{0}^{t}b(s,Z(s))ds$, with \\$b(s,x)=\integ{\R^p}{}y\; K_s(x,dy)=\sum_l l\alpha_l(s,x).$}
\\
2. $C=(C(t)) $ is the quadratic variation of the continuous martingale part of $Z(t)$, $C(t)=(C_{i,j}(t))_{1\leq i,j \leq p}$. For a pure jump process, $C(t)\equiv 0$ .\\
3. \cl{$\nu_t$ is the compensator of the jumps random measure of $(Z_t)$, $\nu_t(dt,dy)= dt\; K_t(Z(t),dy)= dt\;\som{l}{}\alpha_l(t,Z(t))\delta_l(y)$.} \\
4. \cl{The quadratic variation of the $p$-dimensional martingale $M(t)=Z(t)-B(t)$ is, for $1\leq i,j \leq p$,\\
$[M_{ij}](t)= \integ{0}{t} m_{ij}(s) ds$ with $m_{ij}(s)=\integ{\R^p}{}\;y_i\;y_j\; K_s(Z(s),dy)= \sum_{l}l_i\;l_j \alpha_l(s,Z(s))ds. $}\\

\noindent
\cl{Consider now the sequence of normalized pure jump processes  $Z_N(t)=\frac{Z(t)}{N}$} indexed by $N$. The state space of \cl{$Z_N$} is $E_N=\{0,\frac{1}{N},..,1\}^p$,  its transition kernels are $\cl{K^{N}_t(x,dy)= \som{l\in E^-}{}\alpha_l(t,Nx)}\delta_{\frac{l}{N}}(dy)$. Hence, its  
characteristics are \cl{$(B^N,C^N\nu_t^N)$ with \\
1. $B^N(t)= \integ{0}{t} b^N(s,Z_N(s))ds$, with $b^N(s,x)= \som{l}{}\alpha_l(s,Nx)\frac{l}{N}$}\\
2. $C^N(t) \equiv 0$,\\
3.\cl{ $\nu^{N}_{t}(dt,dy)=dt\; K^N_t(Z_N(t),dy)= dt\;\som{l}{}\alpha_l(t,NZ_N(t))\delta_{\frac{l}{N}}(y)$,\\
4. $[M_{i,j}^N](t)= \integ{0}{t} m^N_{ij}(s) ds $ }, with $m^N_{ij}(s)=\integ{\R^p}{}\;y_i\;y_j\; K^N_s(Z(s),dy) =\som{l}{}\alpha_l(s,NZ_N(s))\frac{l_i}{N}\frac{l_j}{N}$.\\

\noindent
\cl{Under (H1), (H2), recall that
$b(t,x)=\som{l\in E^-}{}l\beta_l(t,x)$ and $ x_{x_0}(t) =x_0 +\int_0^t b(s,x_{x_0}(s)) ds.$}\\
We first prove the convergence of the process \cl{$(Z_N(t))$ to} $x_{x_0}(t)$ (which has characteristics $(\integ{0}{t}b(s,x_{x_0}(s))ds,0,0)$) by applying Theorem 3.27 of Chapter IX in \cite{jac87}.
We have to check the following conditions:
\\
(i) $\forall t\in[0,T]$, $\super{0\leq s \leq t}\cl{\norm{B^{N}(t)-\integ{0}{t}b(s,x_{x_0}(s))ds}}\tend{N}{\infty}0$,\\
(ii) $\forall t\in[0,T]$, $[M^N](t) \rightarrow 0$ in probability, \\
(iii) for all $\eta>0$, $\underset{a\rightarrow +\infty}{lim}\underset{N}{limsup}\,\P\left\{ \integ{0}{t}ds\integ{\R^p}{}\norm{y}^2 1_{\norm{y}>a} (y) K^N_s(Z_N(s),dy)>\eta\right\}=0$,\\
(iv) $\forall t\in[0,T]$, 
 $\integ{0}{t}ds\integ{\R^p}{}y \;K^N_s(Z_N(s),dy)\tend{N}{\infty} 0$ in probability.\\
(v) $Z_N(0)\tend{N}{\infty}x_0$ a.s.\\
 Using (H1t), we obtain the uniform convergence of \cl{
$b^N(t,x)\tend{N}{\infty}b(t,x)$ and\\ $[M^N_{ij}](t)\tend{N}{\infty}0$} on $[0,T]\times [0,1]^p$, which ensures conditions (i) and (ii). Condition (v) is satisfied by assumption.  \cl{Since $\integ{\R^p}{}\norm{y}^2 K^N_s(x,dy) <\infty $,} 
(iii) is satisfied. 
\cl{Using now that $\integ{\R^p}{}\norm{y} K^N_s(x,dy) <\infty $ yields (iv)}.
Therefore, $\cl{Z_{N}}(t) \rightarrow x_{x_0}(t)$ in distribution. 
Noting that $b^N(t,x)$ and $[M](t)$ converge uniformly towards $b(t,x)$ and $0$ respectively, and using that the Skorokhod convergence coincides with the uniform convergence when the limit is continuous,
 we get 
\begin{equation}\label{cu:z-det}\super{t\in[0,T]}\norm{Z_{N}(s)-x_{x_0}(s)}
\tend{N}{ \infty } 0 \mbox{ in probability.}
\end{equation}


\noindent
\cl{It remains to study  the process
$Y_N (t)=\sqrt{N}\left(Z_{N}(t)-x_{x_0} (t)\right)$.\\ 
For sake of clarity, we omit in the sequel  the index $x_0$ in $x_{x_0}(t)$.} 
\cl{The jumps of $Y_N$ have size $l/\sqrt{N}$, the  
transition kernel of the jumps random measure is  ${\tilde K}_t^N(y,du)= \som{l}{} \alpha_l(t, Nx(t)+\sqrt{N}y) \delta_{\frac{l}{\sqrt{N}}}(u)$,
$Y_N$ is  a semimartingale with characteristics $({\tilde B}^N,{\tilde C}^N, {\tilde \nu}_t^N)$}\\ 
\cl{
1. $\tilde{B}^N(t)= \integ{0}{t}{\tilde b}^N(s,Y_N(s)) ds$, with\\ 
$ {\tilde b}^N(s,y) = \integ{\R^p}{} u\; {\tilde K}_s^N(y,du)-\sqrt{N}\; b(s,x(s))=
\som{l}{} \alpha_l(s, Nx(s)+\sqrt{N}y) \frac{l}{\sqrt{N}}- \sqrt{N}\; b(s,x(s)).$}
 \\ 
2. \cl{ $ {\tilde C}^N(t) =0$. \\
3. ${\tilde \nu}^N_t(dt,du)=dt\; \tilde{K}^N_t(Y_N(t),du)=dt\; 
(\som{l}{}
\alpha_l(s,Nx(t)+\sqrt{N}Y_N(t)) \delta_{\frac{l}{\sqrt{N}}}(u)). $\\
4. $[\tilde M^N_{ij}](t)= \integ{0}{t} \tilde{m}^N_{ij}(s) ds $, 
with $\tilde{m}^N_{ij}(s)=\integ{\R^p}{}\;u_i\;u_j\; \tilde{K}^N_s(Y_N(s),du) 
=\som{l}{}\alpha_l(s,Nx(s)+ \sqrt{N}Y_N(s))\frac{l_i}{\sqrt{N}}\frac{l_j}{\sqrt{N}}$.}\\

\noindent
\cl{Let us first study $\tilde{B}^N(t)$. Using (H1) $\frac{1}{N}\alpha_l(t, N(x(t)+\frac{y}{\sqrt{N}}))= \beta_l(t, x(t)+\frac{y}{\sqrt{N}})+ r_N(t)$, with $r_N(t)\rightarrow 0$}, \cl{by (H2), $\beta_l(t,.)$ is differentiable and
expanding $\beta_l(t,.)$ around $x(t)$ yields}\\
\cl{$\beta_l(t,x(t)+\frac{y}{\sqrt{N}})= \beta_l(t,x(t))+\som{1}{p} \frac{y_i}{\sqrt{N}}
\frac{\partial \beta_l}{\partial x_i}(t, x(t)) +\;\frac{1}{\sqrt{N}} r'_N(t),$ with 
$r'_N(t)\rightarrow 0$. 
Therefore
 $\frac{1}{\sqrt{N}}\alpha_l(t, N(x(t)+\frac{y}{\sqrt{N}}))= 
\sqrt{N}\beta_l(t, x(t))+\som{i=1}{p} 
y_i\frac{\partial \beta_l}{\partial x_i}(t, x(t))+\sqrt{N}r_N(t)+ r'_N(t). $\\
Hence, we need the additional assumption:\\
(H1t')  $\sqrt{N}(\frac{1}{N}\alpha_l(t, Nx)-\beta_l(t,x))\rightarrow 0$ uniformly w.r.t. $(t,x) \in [0,T]\times [0,1]^p$ as $N\rightarrow \infty$.\\
Then, $\tilde{b}^N(t,y) \rightarrow \som{i=1}{p} 
y_i \som{l}{}l\frac{\partial \beta_l}{\partial x_i}(t,x(t))= \som{i=1}{p}y_i\frac{\partial b}{\partial x_i}(t,x(t))$ and\\ $\tilde{B}^N(t)= 
\integ{0}{t}\som{i=1}{p}\frac{\partial b}{\partial x_i}(s,x(s)) Y_N(s) ds.$\\ 
Therefore, $[\tilde M^N_{ij}](t) \rightarrow \integ{0}{t}\Sigma_{ij}(s,x(s)) ds.$
Checking conditions (iii),(iv),(v) is straightforward. Finally, we obtain that $Y_N$
converges in distribution to the process $Y(t)$ with continuous sample paths, predictable process $\integ{0}{t}\nabla b(s,x(s)) Y(s)ds $ and quadratic variation $\integ{0}{t}\Sigma_{ij}(x(s)) ds$. This  is the diffusion process satisfying the SDE, 
$$dY(t)= (\frac{\partial b_i}{\partial x_j})_{i,j}(t,x(t)) Y(t) dt 
+ \sigma (t,x(t))dB(t)\;;\;Y(0)=0,$$
where $\sigma()$ satisfies
$\sigma(t,x)\trans{\sigma(t,x)}= \Sigma(t,x)$ and $B(t)$ is a $p$-dimensinal Brownian motion. This is an Ornstein-Uhlenbeck type SDE, which can be solved explicitely, leading to  the Gaussian process $G(t)$  previously introduced.} 


\subsection{Extending the contrast approach (for non autonomous diffusion processes and for non constant sampling intervals)}
\label{App:Non-autonomous}
Here, we provide the main line for the extension of the results in \cite{guy12} for non autonomous diffusions and non constant sampling intervals. The complete proof is omitted.
The main point of the proof of Proposition 3.2 in \cite{guy12} relies on the relations (3.7) and (3.8) $\frac{1}{\epsilon\sqrt{\Delta}}A_k(\theta_0)\tend{\epsilon}{0}V_k^{\theta_0}$ and $\frac{1}{\Delta}\deriv{A_k(\theta)}{\theta_i}\tend{\epsilon}{0}D_{k,i}(\theta_0)$. The proof of these relations is based on Taylor stochastic expansion  and the fundamental relation of our contrast approach \eqref{eqg2} . The Taylor stochastic
 expansion of the diffusion was considered in
\cite{fre84} only for autonomous models, but has been extended for time dependent processes by \cite{aze82} and consequently holds when $b$ and
$\Sigma$ are time dependent. Relation \eqref{eqg2} is supported in the autonomous case by the
semi-group property of function $\Phi_\theta$ which leads to an associated
analytic expression of $g_\theta(t)=\integ{0}{t}\Phi_\theta(t,s)\sigma(\theta,s)dB_s$. Since the
semi-group property is stated for non-autonomous cases in \cite{car71}, the extension is immediate.\\ For non constant sampling interval, the simple fact
that relation \eqref{eqg2} holds for any sequence $t_0<t_1<\dots<t_n$ ensures
that the results of Proposition 3.2 in \cite{guy12} hold.\\


\begin{figure}[ht]
\includegraphics[width=0.98\textwidth]{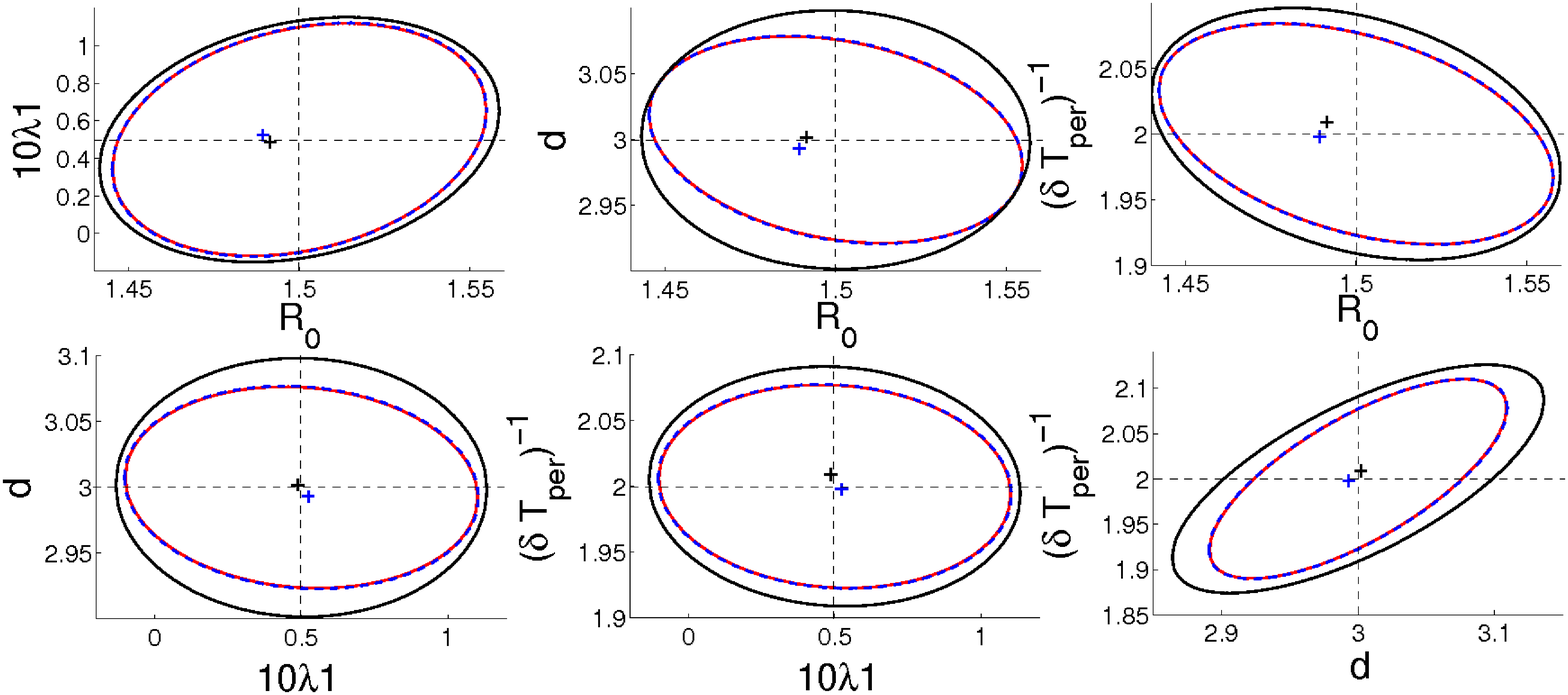}\\
\unnumberedcaption{{\bf Fig. S1} \ev{Point estimators (+) built as averages over 1000 independent simulated trajectories and their associated theoretical confidence ellipsoids  for the $SIRS$ model with demography and seasonal forcing in transmission: $CE$ for 1 obs/day (blue) and for $n=1$ obs/week (black) for $(R_0,d,\lambda_1,\delta)=(1.5,3,0.05,2)$, $T=20$ years and $N=10^7.$ Asymptotic confidence ellipsoid ($n \rightarrow \infty$) is also represented (red). True parameter values are located at the intersection of horizontal and vertical dotted lines.}}
\end{figure}
\begin{figure}[ht]
\includegraphics[width=0.98\textwidth]{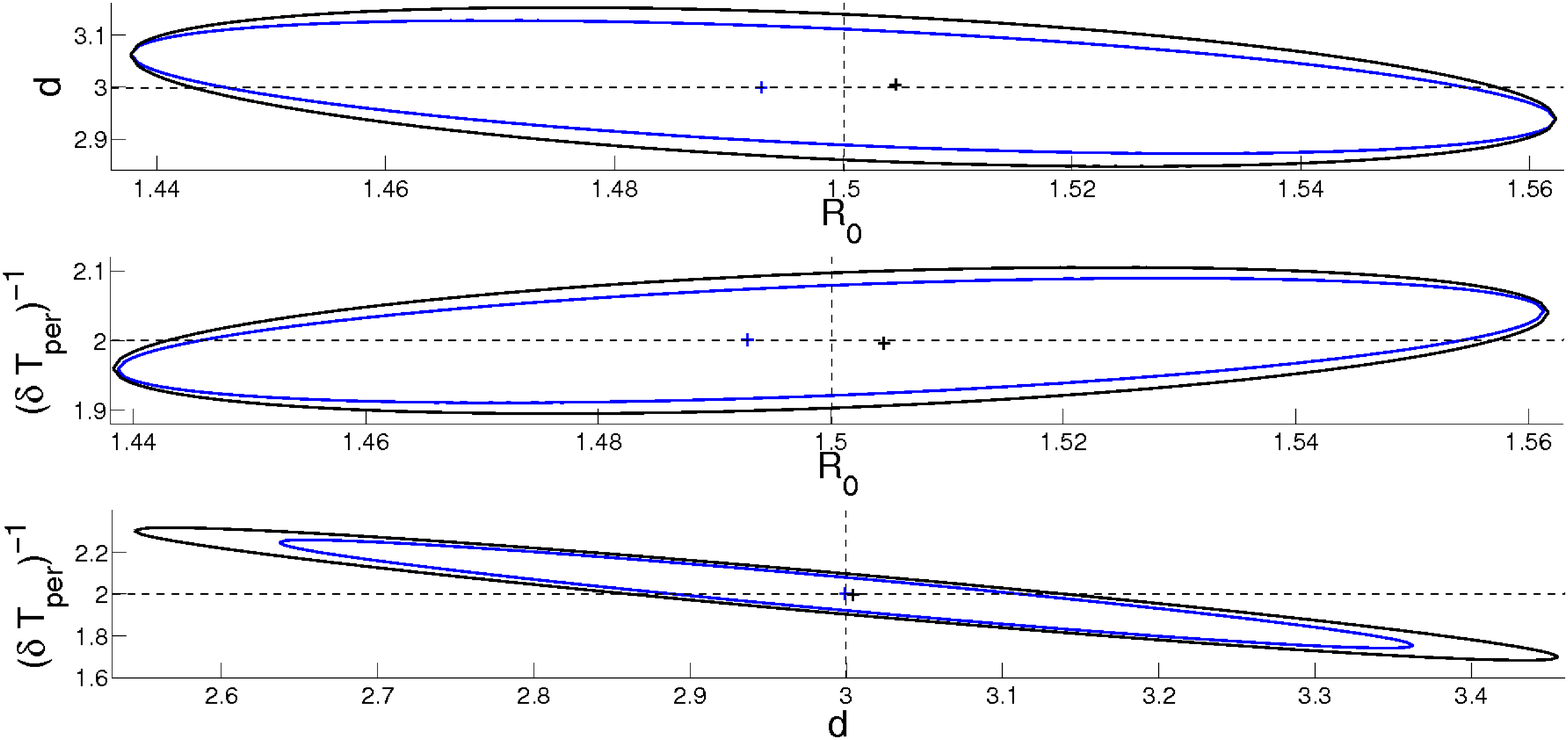}\\
\unnumberedcaption{{\bf Fig. S2} \ev{Point estimators (+) built as averages over 1000 independent simulated trajectories and their associated theoretical confidence ellipsoids  for the $SIRS$ model with demography and no seasonal forcing  ($\lambda_1=0$) in transmission: $CE$ for 1 obs/day (blue) and for $n=1$ obs/week (black) for $(R_0,d,\delta)=(1.5,3,2)$, $T=20$ years and $N=10^7.$ Asymptotic confidence ellipsoid ($n \rightarrow \infty$) is also represented (red). True parameter values are located at the intersection of horizontal and vertical dotted lines.}}
\end{figure}

\FloatBarrier
\section{Acknowledgments}
Partial financial support for this research was provided by Ile de France
Regional Council under MIDEM project in the framework DIM Malinf, and by French Research
Agency, program Investments for the future, project ANR-10-BINF-07\\ (MIHMES). 
\FloatBarrier
\bibliographystyle{spbasic.bst}
\bibliography{biblio}

\end{document}